\documentclass[12pt]{article}

\usepackage{graphicx}	
\usepackage{amsbsy}		
\usepackage{amsfonts}		
\usepackage{amsmath}		
\usepackage{amssymb}		
\usepackage{wasysym}
\usepackage{multirow}
\usepackage{psfrag}
\usepackage{url}
\usepackage{cite}

\newcommand{\CLb}{$CL_b$ }
\newcommand{\CLs}{$CL_s$ }
\newcommand{\CLsb}{$CL_{s+b}$ }

\newcommand{\Bsmumu}{\ensuremath{B_s\to \mu^+\mu^-} }

\newcommand{\figref}[1]{Fig.~\ref{#1}}

\newcommand{\bibref}[1]{Ref.~\cite{#1}}
\newcommand{\secref}[1]{section~\ref{#1}}
\newcommand{\pdf}{{\it p.d.f} }

\newcommand{\invfb}{$\mbox{\,fb}^{-1}$ }

\begin{document}

\hfill {\tt CERN-PH-TH/2011-206}

\hfill {\tt SHEP-11-22}

\def\thefootnote{\fnsymbol{footnote}}

\begin{center}
\Large\bf\boldmath
\vspace*{1.cm} 
The decay $B_s \to \mu^+\mu^-$: updated SUSY constraints\\ and prospects
\unboldmath
\end{center}
\vspace{0.6cm}
\begin{center}
A.G. Akeroyd$^{1,}$\footnote{Electronic address: a.g.akeroyd@soton.ac.uk},
F.~Mahmoudi$^{2,3,}$\footnote{Electronic address: mahmoudi@in2p3.fr} and 
D. Mart{\'{\i}}nez Santos$^{2,}$\footnote{Electronic address: Diego.Martinez.Santos@cern.ch}\\[0.4cm] 
\vspace{0.6cm}
{\sl $^1$ School of Physics and Astronomy, University of Southampton,\\
Highfield, Southampton SO17 1BJ, United Kingdom}\\[0.4cm]
{\sl $^2$ CERN, Physics Department, CH-1211 Geneva 23, Switzerland}\\[0.4cm]
{\sl $^3$ Clermont Universit{\'e}, Universit\'e Blaise Pascal, CNRS/IN2P3,\\
LPC, BP 10448, 63000 Clermont-Ferrand, France}
\end{center}

\renewcommand{\thefootnote}{\arabic{footnote}}
\setcounter{footnote}{0}

\vspace{0.6cm}
\begin{abstract}
We perform a study of the impact of the recently released limits on BR($B_s\to \mu^+\mu^-$) by LHCb and CMS 
on several SUSY models. We show that the obtained constraints can be superior to those which are derived from 
direct searches for SUSY particles in some scenarios, and the use of a double ratio of purely leptonic decays 
involving $B_s\to \mu^+\mu^-$ can further strengthen such constraints. 
We also discuss the experimental sensitivity and prospects for observation of
$B_s\to \mu^+\mu^-$ during the $\sqrt s=7$ TeV run of the LHC, and its potential implications.
\end{abstract}

\newpage
\section{Introduction}
\label{sec:intro}

Rare decays of beauty mesons ($B^\pm$ and $B^0$) are sensitive probes of new particles which arise in models beyond the Standard Model (SM).
In the context of supersymmetric (SUSY) extensions of the SM, the measurements of decays such as $b\to s \gamma$, $B^\pm \to \tau \nu$, and $B^\pm \to D\tau\nu$ provide important constraints on the masses of new particles which are too heavy to be produced directly. Some recent analyses showing constraints on the parameter space of the Minimal Supersymmetric Standard Model (MSSM) 
can be found in \cite{Carena:2006ai,Ellis:2007fu,Mahmoudi:2007gd,Heinemeyer:2008fb,Eriksson:2008cx,Alok:2009wk}. 
Of much interest for the LHC experiments is the unobserved decay $B_s\to \mu^+\mu^-$.
Due its distinct signature, this decay can be searched for by three LHC collaborations: LHCb, CMS and ATLAS.
As pointed out in \cite{Choudhury:1998ze,Babu:1999hn,Huang:2000sm}, $B_s\to \mu^+\mu^-$ is a very effective probe of SUSY models with large ($> 30$) $\tan\beta$, and 
its importance has been emphasised in numerous studies.
The upper limit on the branching ratio (BR) of $B_s\to \mu^+\mu^-$  has been steadily reduced during Run II at the Fermilab Tevatron. 
As of the year 2010, limits of the order of BR($B_s\to \mu^+\mu^-) < 5\times 10^{-8}$ ({\it i.e.} an order of magnitude above the prediction of the SM) were 
obtained by both the CDF \cite{:2007kv} and D0 \cite{Abazov:2010fs} collaborations. 

Recently, the CDF collaboration announced a possible first signal \cite{Aaltonen:2011fi}, although with a low significance. This result has not been confirmed by 
the recent searches at LHCb \cite{LHCb_alone} and CMS \cite{Chatrchyan:2011kr}.
These improved limits for BR($B_s\to \mu^+\mu^-$) further constrain the SUSY parameter space, and we show in this paper that such constraints can be superior to those which are obtained from direct searches for squarks and gluinos.
Using a combination of the individual limits on BR($B_s\to \mu^+\mu^-$) from LHCb and CMS \cite{CMS_plus_LHCb}, 
we present an updated study of the constraints in the context of five distinct SUSY models.
Our numerical analysis is performed with SuperIso v3.2 \cite{Mahmoudi:2007vz,Mahmoudi:2008tp,Mahmoudi:2009zz}, and
we study the following SUSY models: the constrained MSSM (CMSSM), non-universal Higgs mass (NUHM), 
anomaly mediated supersymmetry breaking (AMSB) and gauge mediated supersymmetry breaking (GMSB), all in the context of the MSSM; 
we also study a semi-constrained version of next-to-MSSM (NMSSM).
Moreover, we consider an alternative observable which includes BR($B_s\to \mu^+\mu^-$),
namely, a double ratio of leptonic decays \cite{Grinstein:1993ys,Ligeti:2003hp}.
The double ratio has no dependence on the absolute value of the decay constant $f_{B_s}$, which is the main source of uncertainty in the 
SM prediction for BR($B_s\to \mu^+\mu^-)$, and it was shown in \cite{Akeroyd:2010qy} that this observable can provide competitive 
(or even superior) constraints on the SUSY parameter space.
The main uncertainty in the SM prediction for the double ratio arises from the CKM matrix element $|V_{ub}|$, for which the prospects of precise measurements
at high-luminosity $B$ factories are very promising.
The final integrated luminosity of the operation of the LHC at $\sqrt s =7$ TeV is likely to be
significantly larger than the amount which was anticipated at the start of the run, which could enable the SM prediction
for BR($B_s\to \mu^+\mu^-$) to be probed. We discuss the expected sensitivity to 
BR($B_s\to \mu^+\mu^-$) as a function of the integrated luminosity, as well as the prospects for a measurement 
of a SM-like BR($B_s\to \mu^+\mu^-$) during the 7 TeV run, and its potential implications. 

Our work is organised as follows: in sections 2 and 3 we present a theoretical introduction to the
decay $B_s\to \mu^+\mu^-$ and the double ratio respectively; section 4 contains our numerical analysis of the constraints on various SUSY models that are obtained from the recent upper limit on BR($B_s\to \mu^+\mu^-$);
in section 5 the experimental prospects for $B_s\to \mu^+\mu^-$ are discussed,
and conclusions are contained in section 6.

\section{The decay $B_s \to \mu^+\mu^-$}
\label{sec:Bsmumu}

It has been emphasised in many works \cite{Choudhury:1998ze,Babu:1999hn,Huang:2000sm,Ellis:2005sc,Carena:2006ai,Ellis:2007ss,Mahmoudi:2007gd,Golowich:2011cx}
that the decay $B_s \to \mu^+ \mu^-$ is very sensitive to the presence of SUSY particles. 
At large $\tan\beta$, the SUSY contribution to this process is dominated by the exchange of neutral Higgs bosons, 
and very restrictive constraints are obtained on the supersymmetric parameters.
The ${\rm BR}(B_s\to \mu^+\mu^-)$ can be expressed as \cite{Bobeth:2001sq,Bobeth:2001jm,Buras:2002vd,Mahmoudi:2008tp}
\begin{eqnarray}
\label{BRBsmumu}
\mathrm{BR}(B_s \to \mu^+ \mu^-) &=& \frac{G_F^2 \alpha^2}{64 \pi^3} f_{B_s}^2 \tau_{B_s} m_{B_s}^3 |V_{tb}V_{ts}^*|^2 \sqrt{1-\frac{4 m_\mu^2}{m_{B_s}^2}} \\
&\times& \left\{\left(1-\frac{4 m_\mu^2}{m_{B_s}^2}\right) | C_{Q_1} -C'_{Q_1} |^2 + \left | (C_{Q_2} -C'_{Q_2}) + 2 \, (C_{10} -C'_{10}) \frac{m_\mu}{m_{B_s}} \right |^2\right\} \;, \nonumber
\end{eqnarray} 
where the coefficients $C_{Q_1}$, $C_{Q_2}$, and $C_{10}$ parametrize different contributions. Within the SM, $C_{Q_1}$ and $C_{Q_2}$ are negligibly small, whereas the main contribution entering through $C_{10}$ is helicity suppressed. 
In SUSY, both $C_{Q_1}$ and $C_{Q_2}$ can receive large contributions from scalar exchange,
which was first pointed out (in the context of a different decay, $b\to s l^+l^-$) in \cite{Huang:1998vb}.
 The explicit expressions for the different coefficients can be found in e.g. \cite{Mahmoudi:2008tp}.

The $B_s$ decay constant, $f_{B_s}$, constitutes the main source of uncertainty in  ${\rm BR}(B_s\to \mu^+\mu^-)$.
As of the year 2009 there were two unquenched lattice QCD calculations of $f_{B_s}$, 
by the HPQCD collaboration \cite{Gamiz:2009ku} and FNAL/MILC \cite{Bernard:2009wr} respectively,
which when averaged gave the value $f_{B_s}=238.8\pm 9.5$ MeV \cite{Laiho:2009eu}.
The calculation of \cite{Bernard:2009wr} was updated in \cite{Simone:2010zz}, which gave rise to a higher world average
of $f_{B_s}=250\pm 12$ MeV in the year 2010. Recently, the ETM collaboration announced its result of $f_{B_s}=232\pm 10$ MeV \cite{:2011gx}.
At the Lattice 2011 conference \cite{Lattice2011}, new results 
by FNAL/MILC ($f_{B_s}=242\pm 9$ MeV) and the HPQCD collaboration ($f_{B_s}=226\pm 10$ MeV 
\cite{Shigemitsu:2011sp}
and $f_{B_s}=225\pm 4$ MeV \cite{McNeile:2011ng}) 
suggest that an updated world average would be lower than that of the year 2009.
In our numerical analysis we will use $f_{B_s}=238.8\pm 9.5$ MeV \cite{Laiho:2009eu}.

To study the constraints on the parameter spaces of SUSY scenarios, we use the newly released combined limit from LHCb and CMS at 95\% C.L. \cite{CMS_plus_LHCb}:
\begin{equation}
\mathrm{BR}(B_s\to\mu^+\mu^-) < 1.1 \times 10^{-8}\;. \label{BRBsmumu_limit}
\end{equation}
More details are given in section~\ref{sec:prospects}. In order to take into account the theoretical uncertainties, in our numerical analysis we will use the following limit
\begin{equation}
\mathrm{BR}(B_s\to\mu^+\mu^-) < 1.26 \times 10^{-8}
\end{equation}
to constrain the parameter spaces of the SUSY models under consideration.

\section{The double ratios of purely leptonic decays}
\label{sec:doubleratio}

The main uncertainty in the theoretical prediction of $B_s\to\mu^+\mu^-$ is $f_{B_s}$. 
As described in Section \ref{sec:Bsmumu},
$f_{B_s}$ is now being evaluated in the unquenched approximation by various lattice
collaborations. The error (which is currently around $5\%$ or less) 
has been reduced over time, and 
the central values of $f_{B_s}$ from the various collaborations are in reasonable agreement. 
The prospects for a further reduction of the error are good.
However, despite the continuing improvement in the calculations of $f_{B_s}$
our view is that it is instructive to consider other observables which
involve $B_s\to\mu^+\mu^-$ but do not depend on the decay constants,
and to compare the
constraints on the SUSY parameter space with those which are obtained 
from $\mathrm{BR}(B_s\to\mu^+\mu^-)$ alone. 
One such observable which involves $B_s\to\mu^+\mu^-$, but has essentially no dependence on the
absolute values of the decay constants, is a double ratio involving the leptonic decays
$B_u\to \tau\nu, B_s\to \mu^+\mu^-, D\to \mu\nu$ and $D_s\to \mu\nu/\tau\nu$ \cite{Grinstein:1993ys,Ligeti:2003hp}.

One such double ratio is defined by:
\begin{equation}
\frac{\Gamma(B_s\to \mu^+\mu^-)}{\Gamma(B_u\to\tau\nu)}
\frac{\Gamma(D\to \mu\nu)}{\Gamma(D_s\to\mu\nu)}\sim
\frac{|V_{ts}V_{tb}|^2}{|V_{ub}|^2}\;\frac{\alpha^2}{\pi^2}\;
\frac{(f_D/f_{D_s})^2}{(f_B/f_{B_s})^2} \;.
\label{doub-rat}
\end{equation}

The quantity $(f_B/f_{B_s})/(f_D/f_{D_s})$ deviates from unity by small corrections
of the form $m_s/m_b$ and $m_s/m_c$. The double ratio would be equal to one in the
heavy quark limit of a very large mass for the $b$ and $c$ quarks, and in the limit of exact SU(3) flavour symmetry
($m_s\to 0$).  
A calculation in \cite{Grinstein:1993ys} gives $(f_B/f_{B_s})/(f_D/f_{D_s})=0.967$, and subsequent works
\cite{Oakes:1994tj} also give values very close to 1, with a very small error. 
Unquenched lattice calculations of the ratios $f_{D_s}/f_{D}$ and $f_{B_s}/f_{B}$ have a precision of the
order of $1\%$ (e.g. \cite{Simone:2010zz}), from which it can be inferred that the numerical value of the 
double ratio is very close to 1.
In our numerical analysis we will take $(f_B/f_{B_s})/(f_D/f_{D_s})=1$.  
Importantly, the absolute values of the decay constants do not determine the value of the double ratio,
in contrast to the case of BR($B_s\to \mu^+\mu^-$) alone in Eq.~(\ref{BRBsmumu}).
Instead, $|V_{ub}|$ replaces $f_{B_s}$ as the only major source of uncertainty, as can be seen from Eq.~(\ref{doub-rat}).
Information on $|V_{ub}|$ is available from direct measurements of semileptonic decays of $B$ mesons, both
inclusive ($B\to X_u \ell\nu$) and exclusive ($B\to \pi \ell\nu$). Moroever, global fits \cite{Bona:2009cj}
in the context of the SM give additional experimental information on $|V_{ub}|$.
Due to its different theoretical uncertainties, the double ratio is an
alternative observable which includes $B_s\to \mu^+\mu^-$, and can provide competitive constraints on SUSY parameters.
A comparison of the constraints on specific SUSY models
from the double ratio and from BR($B_s\to \mu^+\mu^-$) alone is
of much interest because the theoretical input parameters $|V_{ub}|$ and $f_{B_s}$ for these two observables are independent.
Such a comparative study was performed for the first time in \cite{Akeroyd:2010qy}, and
it was shown that the double ratio can provide stronger constraints.
In particular, the constraints from the double ratio are maximised (minimised) for smaller (larger) $|V_{ub}|$,
while the constraints from BR$(B_s\to \mu^+\mu^-)$ are maximised (minimised) for larger (smaller) $f_{B_s}$. 

We will perform an updated study of these two observables using the recently improved bounds on BR($B_s\to\mu^+\mu^-$) from the LHCb \cite{LHCb_alone} and CMS
\cite{Chatrchyan:2011kr} collaborations. In the previous study of the double ratio \cite{Akeroyd:2010qy} the value
$|V_{ub}|=(3.92\pm 0.45 \pm 0.09)\times 10^{-3}$ \cite{Asner:2010qj}
was used, which is an average of the exclusive and inclusive determinations of $|V_{ub}|$. We note that this world average
does not include three recent measurements of $|V_{ub}|$, of which two are from the exclusive channel \cite{:2010uj,Ha:2010rf} and one
is from the inclusive channel \cite{Sigamani:2011ne}. The inclusion of these measurements would only have a small effect
on the world average, and so for simplicity we will use $|V_{ub}|=(3.92\pm 0.45 \pm 0.09)\times 10^{-3}$, as done in
\cite{Akeroyd:2010qy}.

We note that the exclusive determination of $|V_{ub}|$ suggests values of $|V_{ub}|$ which are
below the central value of the world average. The exclusive determination of $|V_{ub}|$ requires a theoretical
calculation of one hadronic form factor $f_+(q^2)$ (where $q$ is the momentum of $\ell$).
For $q^2 > 16 \,{\rm GeV}^2$ one can use lattice QCD to calculate $f_+(q^2)$,
 while for  $q^2 < 16 \,{\rm GeV}^2$ non-lattice techniques must be used. In both regions
of $q^2$ the extracted value of $|V_{ub}|$ is below the central value of the world average.
The inclusive determination of $|V_{ub}|$, which does not have a dependence on lattice QCD, 
suggests values which are above the central value of the world average. Prospects for precise measurements
of $|V_{ub}|$ in the inclusive channel are very good at high-luminosity $B$ factories.
In particular, the method used in \cite{:2009tp,Sigamani:2011ne} is a very promising
approach because the theoretical errors are greatly reduced by employing a low cut on the
momentum of the $\ell$ ($p_\ell > 1$ GeV), which keeps $90\%$ of the phase space of $B\to X_u \ell \nu$.
This anticipated experimental improvement in the measurement of $|V_{ub}|$ bodes well for the double ratio
as an alternative observable with which to constrain SUSY.
It is important to emphasise that $f_{B_s}$ 
is currently known with greater precision than $|V_{ub}|$, and this
may also be the case in the era of a high-luminosity B factory. However, we note that the central values of these 
unrelated input parameters plays a major role in determining which observable gives the stronger constraints,
as will be discussed in our numerical analysis.

The double ratio also has the attractive feature of using ongoing measurements of
BR($D_s\to \mu\nu/\tau\nu$) and BR($D\to \mu\nu$). Such decays are not usually discussed when constraining SUSY parameters
(although see \cite{Akeroyd:2009tn} for a discussion of  $D_s\to \mu\nu/\tau\nu$ in this regard), but increased precision
in their measurements would enhance the capability of the double ratio to probe the SUSY parameter space.
The decay $B_u \to \tau \nu$ alone is very sensitive to the presence of a charged Higgs boson ($H^\pm$) and
provides a strong constraint on $\tan\beta$ and the
mass of $H^\pm$ in SUSY models \cite{Hou:1992sy,Akeroyd:2003zr,Itoh:2004ye,Isidori:2006pk}.
The experimental prospects for precise measurements of all the decays in the double ratio are
very promising. The precision in the measurements of  BR($D_s\to \mu\nu$) and BR($D_s\to \tau\nu$) will
be improved at the ongoing BES-III experiment \cite{Asner:2008nq}, and at high-luminosity $B$ factories
operating at a centre-of-mass energy of $\sqrt s\sim 10.6$ GeV (and also possibly at energies in the charm threshold region).
Similar comments apply to the prospects for significantly improved measurements of BR($D\to \mu\nu$) and BR$(B_u \to \tau \nu)$. For 
more details about the calculation of these decays we refer the reader to \cite{Akeroyd:2010qy}.

In this analysis we use
\begin{equation}
 R \equiv \frac{\eta}{\eta_{\mathrm{SM}}} \;, \label{ratio}
\end{equation}
where
\begin{equation}
\eta \equiv \left(\frac{\mathrm{BR}(B_s\to\mu^+\mu^-)}{\mathrm{BR}(B_u \to \tau \nu)}\right) \Big/ \left(\frac{\mathrm{BR}(D_s\to\tau\nu)}{\mathrm{BR}(D\to\mu\nu)}\right) \;. \label{ratio_eta}
\end{equation}

\begin{figure}[!t]
\begin{center}
\includegraphics[width=7.cm]{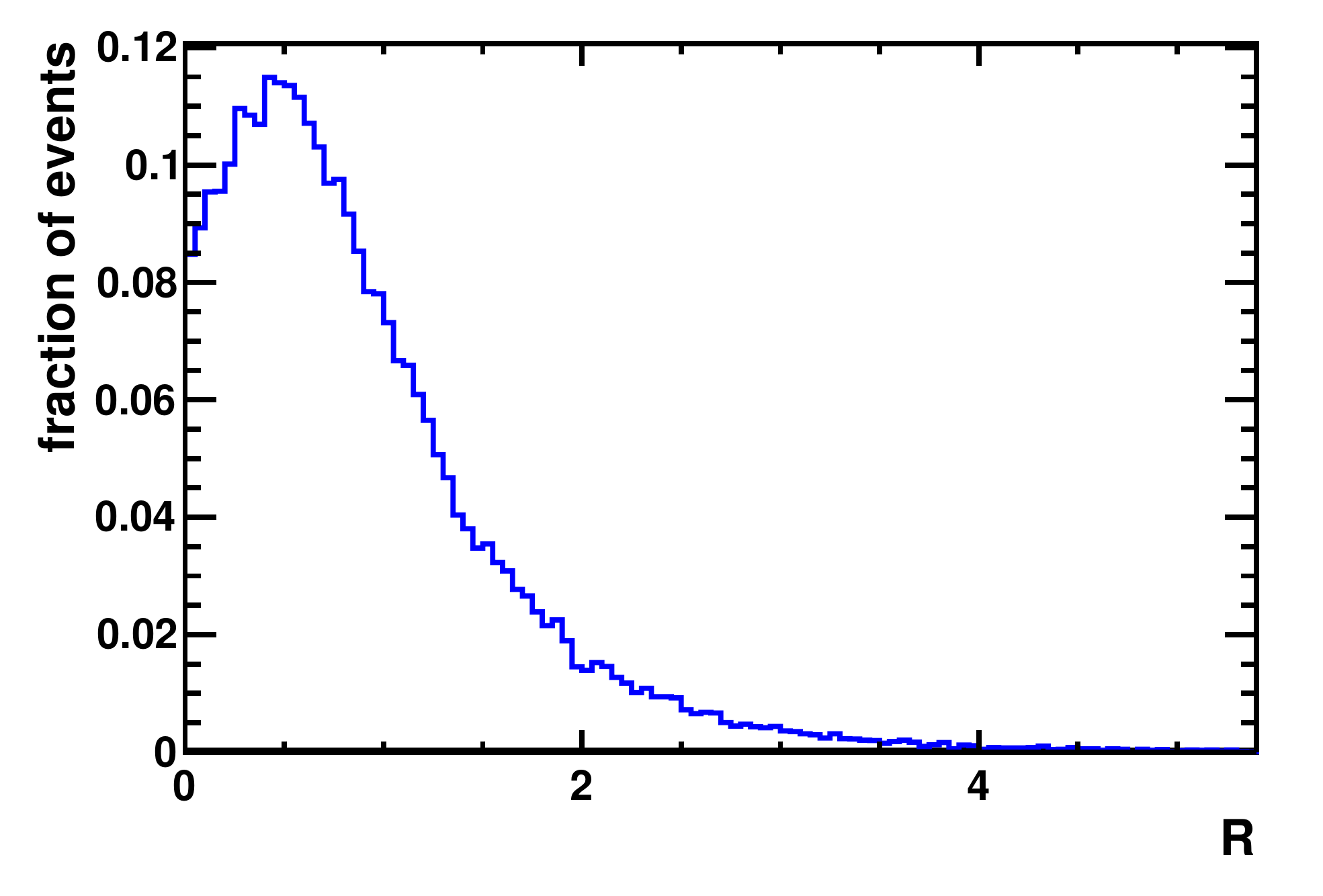}
\caption{Probability distribution function (\pdf) of the double ratio $R$. For \Bsmumu the \pdf  was obtained based on the
C.L. from \cite{CMS_plus_LHCb}. For the other three decays, their measurements are modelled as Gaussians.}
\label{fig:R}
\end{center}
\end{figure}

The theoretical evaluation of $\eta_{\mathrm{SM}}$ gives $( 2.47 \pm 0.58 )\times 10^{-7}$ where the main uncertainty comes from $V_{ub}$. To determine the experimental limit on the ratio $R$, we combine the limits on the individual branching fractions, namely $\mathrm{BR}(B_s\to\mu^+\mu^-)$, $\mathrm{BR}(B_u \to \tau \nu)$, $\mathrm{BR}(D_s\to\tau\nu)$ and $\mathrm{BR}(D\to\mu\nu)$. 
To compute the \pdf of $R$, we use a Gaussian distribution for the measured decays, and a ``truncated'' Gaussian \pdf for the upper limit in \eqref{BRBsmumu_limit}. 
We consider two different approaches. 
The first approach consists in building first the \pdf for $\mathrm{BR}(B_s\to\mu^+\mu^-)$ which reproduces the 90\% and 95\% C.L. experimental limits, and to combine it with the Gaussian \pdf of the other involved decays.
The second approach determines the \pdf from the derivative of the \CLsb with $\mathrm{BR}(B_s\to\mu^+\mu^-)$, which is
extracted from the derivative of \CLs shown in \bibref{CMS_plus_LHCb} and the almost constant behaviour of \CLb. \figref{fig:R} shows the $R$ \pdf.

Both approaches agree and provide the upper limit for $R$, at 95\% C.L.:
\begin{equation}
R < 2.3 \;, 
\label{R_doubrat}
\end{equation}
in which the uncertainty from $V_{ub}$ is taken into account. In our numerical analysis we use
\eqref{R_doubrat} to constrain the supersymmetric parameter space in various scenarios in the MSSM and NMSSM.

\section{Constraints on SUSY Models}
\label{sec:constraints}

We consider five distinct SUSY models in order to illustrate the impact of the new limits on $\mathrm{BR}(B_s\to\mu^+\mu^-)$ on 
the SUSY parameter spaces. All previous studies have been carried out before the  LHCb \cite{LHCb_alone} and CMS \cite{Chatrchyan:2011kr} limits were released\footnote{In an updated version of Ref. \cite{Farina:2011bh} the impact of the latest LHCb \cite{LHCb_alone} and CMS \cite{Chatrchyan:2011kr} limits on $\mathrm{BR}(B_s \to \mu^+\mu^-)$ is studied amongst other observables in a global CMSSM fit.}.
Some very recent works \cite{Dutta:2011bk} study the impact of the latest CDF result \cite{Aaltonen:2011fi} only,
and address the case of the excess of events being a genuine signal.
Moreover, none of the previous studies have considered the double ratio, apart from our earlier work in \cite{Akeroyd:2010qy}
in which two of the five SUSY scenarios were discussed.

For each scenario we also check the constraints from direct searches for 
Higgs bosons and delimit the regions where the lightest supersymmetric particle (LSP) is charged.
All the flavour observables are calculated with the SuperIso v3.2 program \cite{Mahmoudi:2007vz,Mahmoudi:2008tp,Mahmoudi:2009zz}. The spectrum 
of the MSSM points is generated with SOFTSUSY-3.1.7 \cite{Allanach:2001kg} and we used NMSPEC program from the NMSSMTools 3.0.0 package 
\cite{Ellwanger:2006rn} for the NMSSM points. For every generated MSSM point we check if it fulfills the constraints from the Higgs searches 
using HiggsBounds-3.2.0 \cite{Bechtle:2008jh,Bechtle:2011sb}.  
The value of $m_t = 173.3$ GeV \cite{:1900yx} is used throughout.

\subsection{CMSSM}
\label{sec:cmssm}
The first model we consider is the constrained MSSM (CMSSM) \cite{mSUGRA}, 
which is characterized by the set of parameters $\{m_0, m_{1/2}, A_0, \tan\beta, \mbox{sgn}(\mu)\}$. The CMSSM model 
invokes unification boundary conditions at a very high scale $m_{GUT}$ where the universal mass parameters are specified.

To explore the CMSSM parameter space, we generate about 300,000 random points scanning over the ranges 
$m_0 \in [50,2000]$ GeV, $m_{1/2} \in [50,2000]$ GeV, $A_0 \in [-2000,2000]$ GeV and $\tan\beta \in [1,60]$ with positive $\mu$ 
(as favoured by the muon ($g-2$) measurements). 
\begin{figure}[!t]
\begin{center}
\includegraphics[width=8.cm]{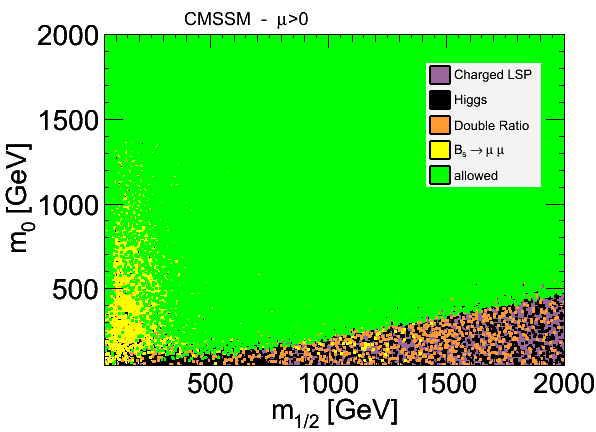}\\[3mm]
\includegraphics[width=8.cm]{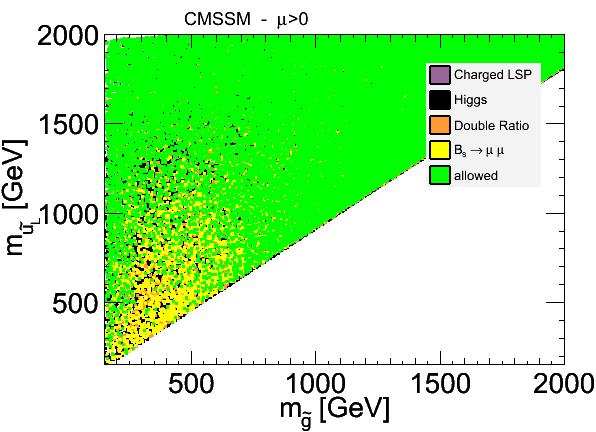}\\[3mm]
~~\includegraphics[width=8.cm]{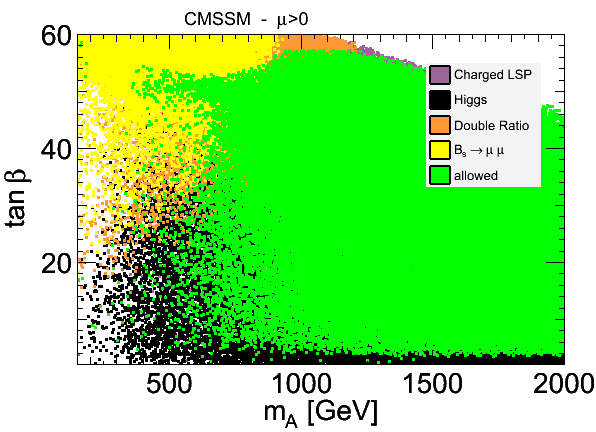}
\caption{Constraints from $\mathrm{BR}(B_s \to \mu^+\mu^-)$ and the double ratio $R$ in the CMSSM planes $(m_{1/2},m_0)$ in the upper panel, $(m_{\tilde g},m_{{\tilde u}_L})$ in the middle panel and $(m_A,\tan\beta)$ in the lower panel. The colour coding is given in the text and the constraints are applied in the order they appear in the legend, with the allowed points in green displayed on top.}
\label{fig:cmssm_mup}
\end{center}
\end{figure}

\begin{figure}[!th]
\begin{center}
\includegraphics[width=8.cm]{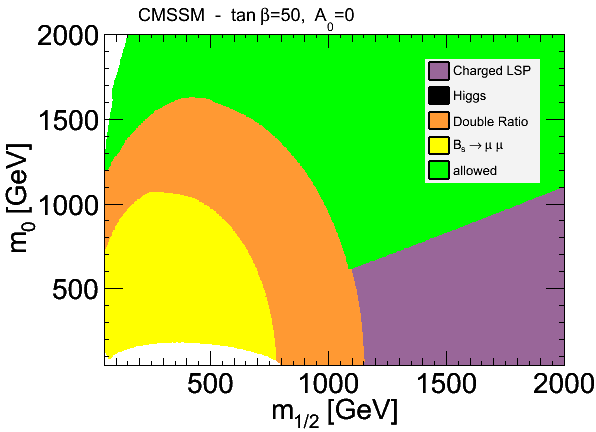}~~~~\includegraphics[width=8.cm]{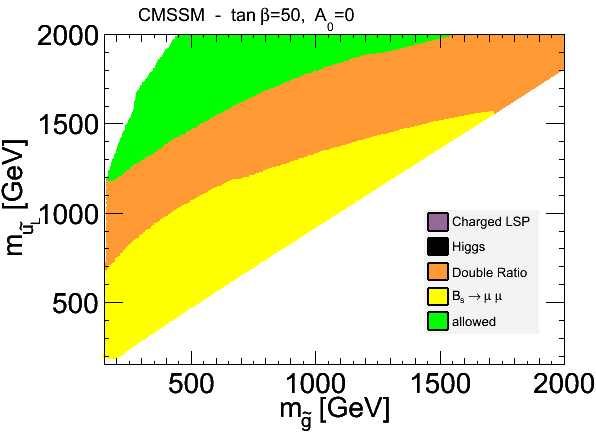}\\[4mm]
\includegraphics[width=8.cm]{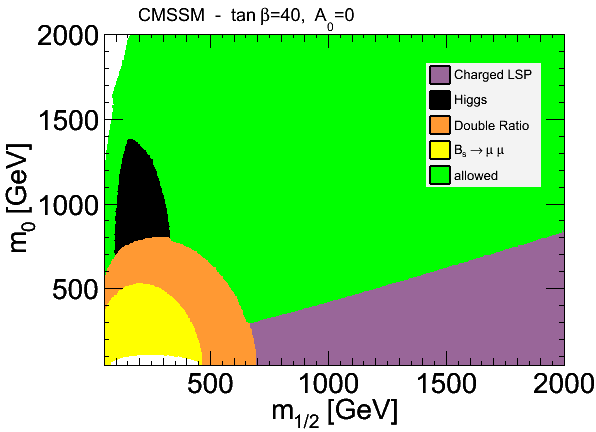}~~~~\includegraphics[width=8.cm]{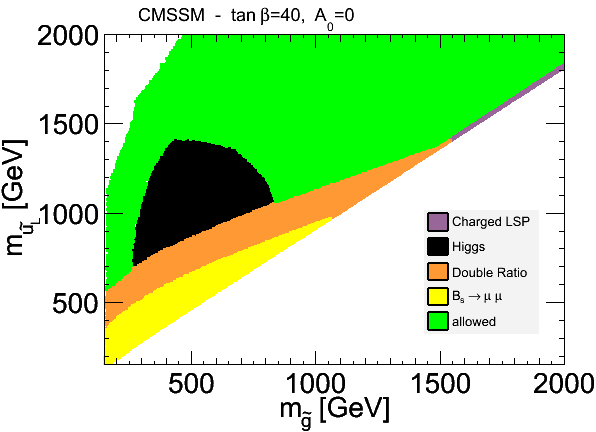}\\[4mm]
\includegraphics[width=8.cm]{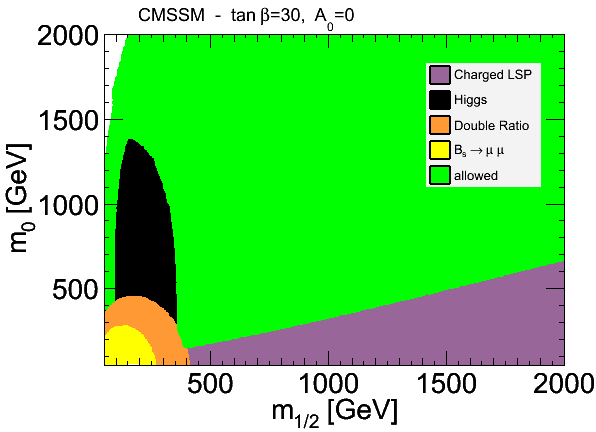}~~~~\includegraphics[width=8.cm]{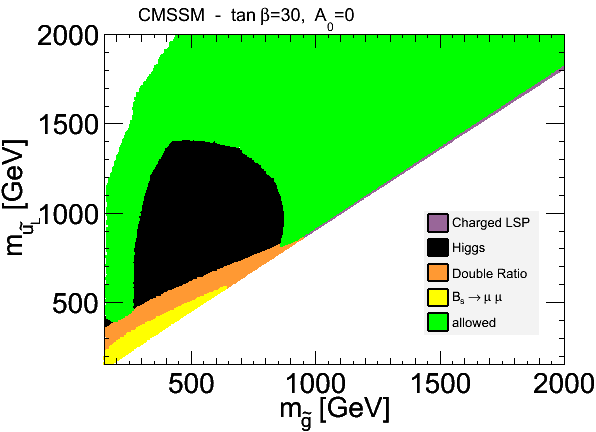}
\caption{Constraints from $\mathrm{BR}(B_s \to \mu^+\mu^-)$ and the double ratio $R$ in the CMSSM planes $(m_{1/2},m_0)$ on the left and $(m_{\tilde{g}},m_{\tilde{u}_L})$ on the right, for $A_0=0$ and $\tan\beta=50$ (upper panel), $\tan\beta=40$ (middle panel) and $\tan\beta=30$ (lower panel).}
\label{fig:cmssm_tb}
\end{center}
\end{figure}

The results are displayed in Fig.~\ref{fig:cmssm_mup}, where the 
four-dimensional space is projected into a plane. When interpreting these results it is therefore important to remember that 
each point in the figures corresponds to a multi-dimensional parameter space in the variables which are not
displayed on the $x$-axis and the $y$-axis. 

In order to show the viable parameter space of the SUSY scenario under investigation,
in all the figures we introduce a colour coding which is applied sequentially.
Areas which are disallowed theoretically are in white. Next, the points which are disallowed phenomenologically
are plotted, which are those with a charged LSP (in violet) and those 
which are excluded by the direct searches for Higgs bosons (in black). In this way, these points lie in the background. 
On top of them, the points excluded by the double ratio $R$ (in orange) are displayed, superseded by the points excluded by 
$\mathrm{BR}(B_s\to\mu^+\mu^-)$ (in yellow). Finally the allowed points (in green) are shown in the foreground. 

These indirect constraints on the CMSSM parameter space from $\mathrm{BR}(B_s\to\mu^+\mu^-)$ are competitive
with the direct constraints from searches for squarks and gluinos by ATLAS and CMS \cite{atlas-cms-EPS}.
As expected, one can see strong constraints on small $m_A$ and large $\tan\beta$ values. At large $\tan\beta$ ($\gtrsim$ 30), these constraints are stronger than those obtained from $\mathrm{BR}(B\to X_s\gamma)$ \cite{Mahmoudi:2007gd}.

In order to better quantify the impact of $\mathrm{BR}(B_s\to\mu^+\mu^-)$ and $R$, we show in Fig.~\ref{fig:cmssm_tb} the constraints for fixed values of $\tan\beta$ (=30, 40 and 50) and $A_0=0$. One striking result here is that the double ratio, being a combination of four different flavour observables, extends impressively the constraints obtained by $\mathrm{BR}(B_s\to\mu^+\mu^-)$ alone, as was pointed out in \cite{Akeroyd:2010qy}. Also, for $\tan\beta=50$, the constraints from the flavour observables go far beyond the direct search limits by the ATLAS and CMS collaborations for the same scenario.

The SUSY contributions to $B_u\to \tau\nu$ gives rise to a scale factor which multiplies BR($B_u\to \tau\nu$). 
When we manually set this scale factor to be equal to 1 (as in the SM), the excluded region of the
plane  $[m_0,m_{1/2}]$ does not change much. Therefore we conclude that the points i) and ii) above are the main reasons why the
double ratio gives the superior constraints.

\subsection{NUHM}
\begin{figure}[!t]
\begin{center}
\includegraphics[width=8.cm]{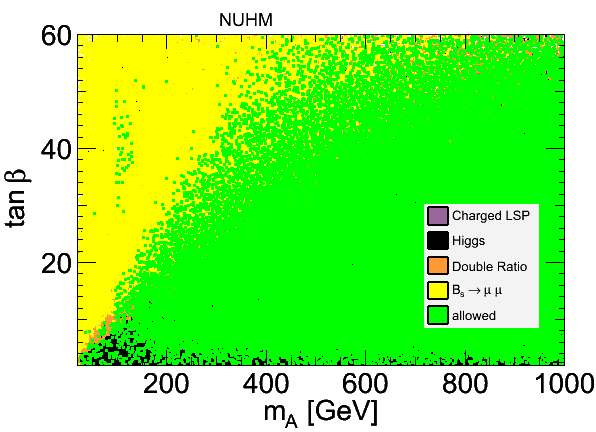}
\caption{Constraints from $\mathrm{BR}(B_s \to \mu^+\mu^-)$ and the double ratio $R$ in the NUHM plane $(m_A,\tan\beta)$. The colour coding is explained in section \ref{sec:cmssm}.}
\label{fig:nuhm}
\end{center}
\end{figure}

The second model we consider involves non-universal Higgs masses (NUHM) \cite{Ellis:2002wv}. This model
generalizes the CMSSM, allowing for the GUT scale mass parameters of the Higgs doublets to have values different from $m_0$, {\it i.e.} $m_{H_1} \neq m_{H_2} \neq m_0$. These two additional parameters with dimension of mass can be traded for two other parameters at a lower scale, 
which can be conveniently chosen as the $\mu$ parameter and the mass $m_A$ of the CP-odd Higgs boson.

We generate about 300,000 random points in the ranges $m_0 \in [50,2000]$ GeV, $m_{1/2} \in [50,2000]$ GeV, $A_0 \in [-2000,2000]$ GeV, $\tan\beta \in [1,60]$, $\mu \in [-2000,2000]$ GeV and $m_A \in [20,1000]$ GeV. The results are presented in Fig.~\ref{fig:nuhm}. Again the constraints are very important, 
and restrict strongly the region of large $\tan\beta$ / small $m_A$.

\begin{figure}[!t]
\begin{center}
\includegraphics[width=8.cm]{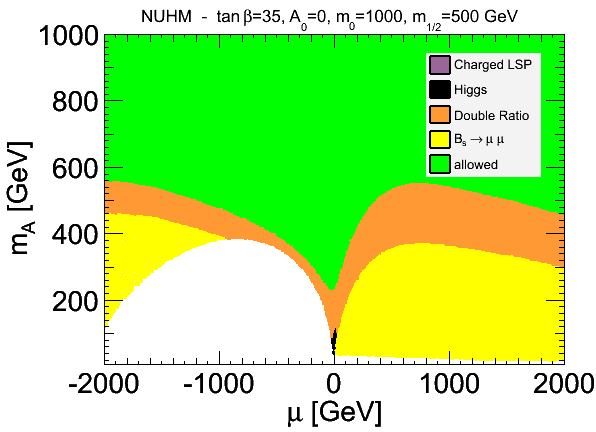}~~~~\includegraphics[width=8.cm]{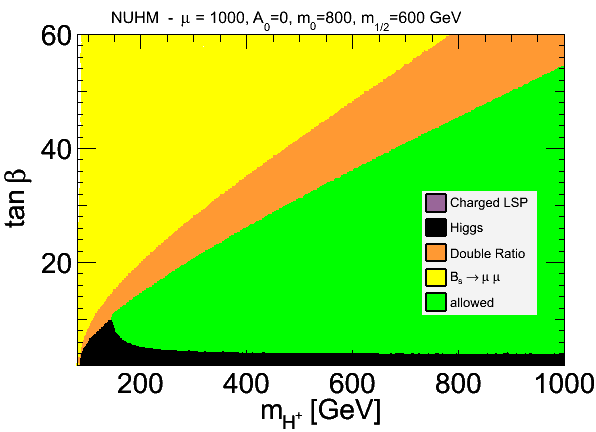}
\caption{Constraints from $\mathrm{BR}(B_s \to \mu^+\mu^-)$ and the double ratio $R$ in the NUHM parameter plane $(\mu,m_A)$ with $\tan\beta = 35$, $A_0=0$, $m_0=1000$ and $m_{1/2}=500$ GeV on the left, and in the plane $(m_{H^+},\tan\beta)$ for $\mu=1000$, $A_0=0$, $m_0=800$ and $m_{1/2}=600$ GeV on the right.}
\label{fig:nuhm2}
\end{center}
\end{figure}

In Fig.~\ref{fig:nuhm2} we show two examples in the two-dimensional parameter planes $(\mu,m_A)$ and $(m_{H^+},\tan\beta)$ with the rest of parameters being fixed. As can be seen from the figures, a large part of the parameter space is restricted by 
$\mathrm{BR}(B_s\to\mu^+\mu^-)$ and $R$ observables, whereas in the same plane one would not get any 
constraints from $\mathrm{BR}(B\to X_s\gamma)$ for $\mu>0$ \cite{Mahmoudi:2007gd}.

\subsection{AMSB}
We can now focus on another supersymmetry breaking scenario, namely the Anomaly Mediated Supersymmetry Breaking (AMSB) \cite{AMSB}.
This is a special case of gravity mediation in which there is no direct tree-level coupling that transmits the SUSY breaking 
in the hidden sector to the visible one. 
The breaking is communicated through the conformal anomaly. 
The free parameters of the minimal model consist of $\{m_0, m_{3/2}, \tan\beta, \mbox{sgn}(\mu)\}$.

 \begin{figure}[!t]
\begin{center}
\includegraphics[width=8.cm]{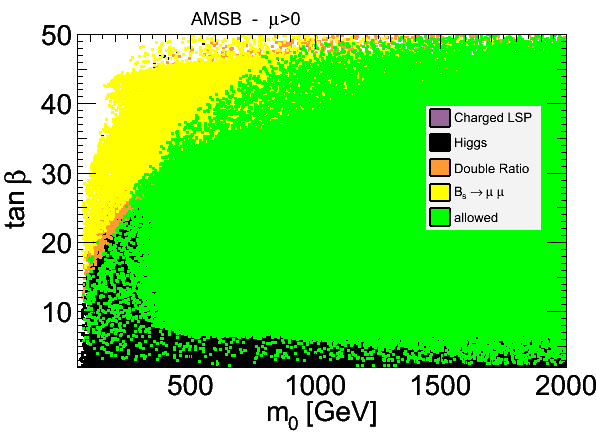}~~~~\includegraphics[width=8.cm]{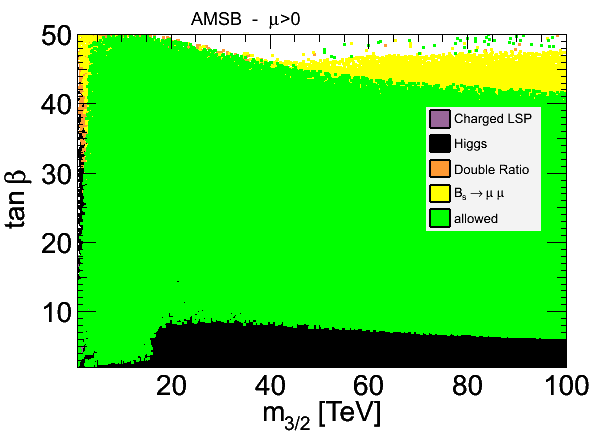}
\caption{Constraints from $\mathrm{BR}(B_s \to \mu^+\mu^-)$ and the double ratio $R$ in AMSB. The colour coding is given in section \ref{sec:cmssm}.}
\label{fig:amsb_mup}
\end{center}
\end{figure}

\begin{figure}[!t]
\begin{center}
\includegraphics[width=8.cm]{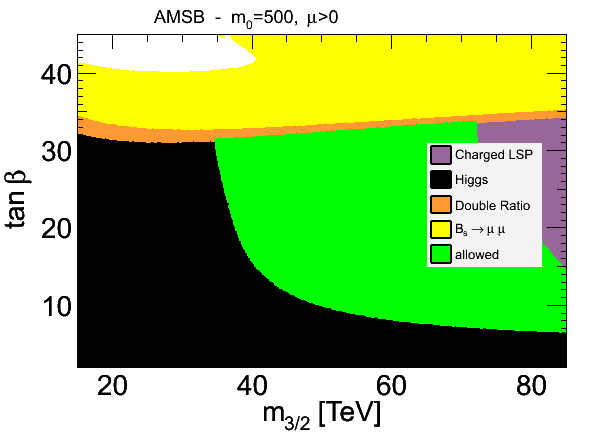}~~~~\includegraphics[width=8.cm]{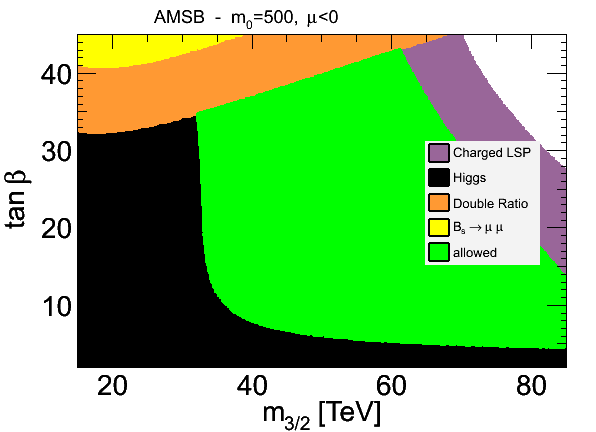}
\caption{Constraints from $\mathrm{BR}(B_s \to \mu^+\mu^-)$ and the double ratio $R$ in the AMSB parameter plane $(m_{3/2},\tan\beta)$ for $m_0=500$ GeV and $\mu>0$ (on the left) and $\mu<0$ (on the right).}
\label{fig:amsb_m0}
\end{center}
\end{figure}

\begin{figure}[!t]
\begin{center}
\includegraphics[width=8.cm]{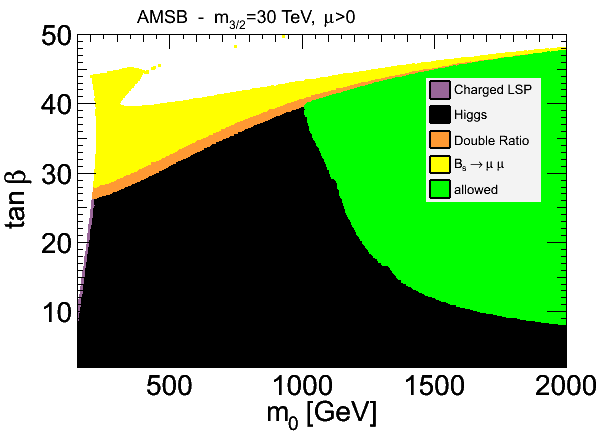}~~~~\includegraphics[width=8.cm]{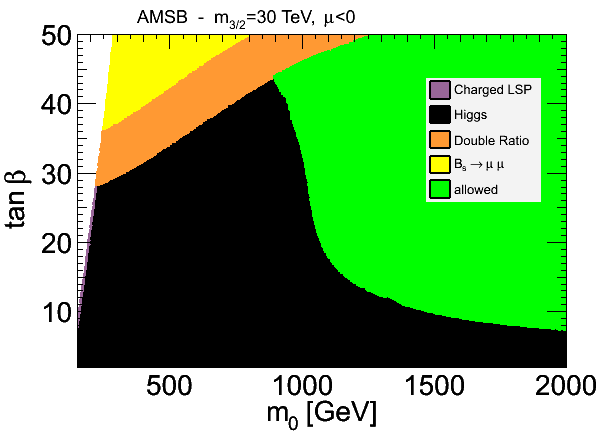}
\caption{Constraints from $\mathrm{BR}(B_s \to \mu^+\mu^-)$ and the double ratio $R$ in the AMSB parameter plane $(m_0,\tan\beta)$ for $m_{3/2}=30$ TeV and $\mu>0$ 
(on the left) and $\mu<0$ (on the right).}
\label{fig:amsb_m32}
\end{center}
\end{figure}

Previous studies were performed in \cite{Heinemeyer:2008fb,Arbey:2011gu}.
To explore the parameter space of AMSB, we scan over $m_0 \in [50,2000]$ GeV, $m_{3/2} \in [1,100]$ TeV and $\tan\beta \in [1,60]$, and generate 300,000 random model points. The results are presented in Fig.~\ref{fig:amsb_mup} and show stronger limits for low values of $m_0$ and large $\tan\beta$.

Fig.~\ref{fig:amsb_m0} and Fig.~\ref{fig:amsb_m32} show the results in two-dimensional planes in order to see better the extent
 of the constraints. In the plane $(m_{3/2},\tan\beta)$ for $m_0=500$ GeV, essentially all of the points with $\tan\beta \gtrsim 30$ are disfavoured regardless of the value of $m_{3/2}$. In the plane $(m_0,\tan\beta)$ with $m_{3/2}=30$ TeV, one obtains 
strong constraints for small $m_0$ / large $\tan\beta$. Scenarios with $\mu<0$ show similar effects, 
with the constraints being less pronounced. It is also evident that a large portion of the parameter space 
is already excluded by the constraints from the direct searches for Higgs bosons, as implemented in the HiggsBounds program.

\subsection{GMSB}
The last MSSM scenario that we consider is the Gauge Mediated Supersymmetry Breaking (GMSB) scenario \cite{GMSB}, which consists of the SUSY breaking sector and the messenger sector. The latter can be taken as a $5 + \bar{5}$ of the SU(5) which contains the Standard Model group, and therefore the gauge coupling unification is not affected. The minimal model is characterized by the set of parameters $\{\Lambda, M_{mess}, N_5, c_{grav}, \tan\beta, \mbox{sgn}(\mu)\}$. For our study, we consider $N_5=1$, $c_{grav}=1$ and generate about 300,000 random points in the ranges $\Lambda \in [10,500]$ TeV, $M_{mess} \in [10^2,10^{14}]$ TeV and $\tan\beta \in [1,60]$ with $\Lambda < M_{mess}$.

\begin{figure}[!th]
\begin{center}
\includegraphics[width=8.cm]{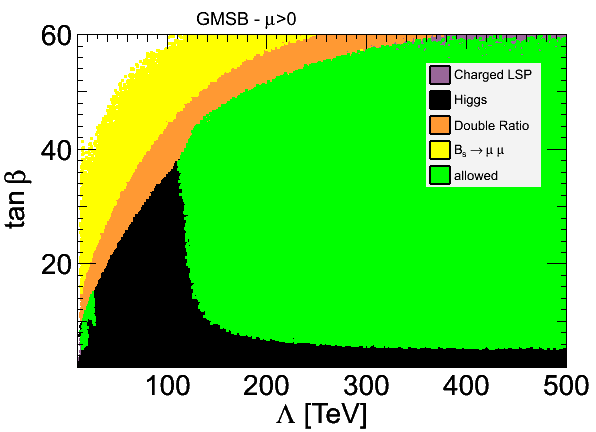}~~~~\includegraphics[width=8.cm]{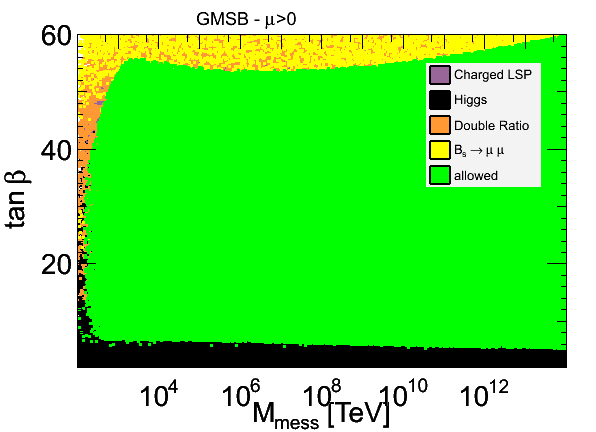}
\caption{Constraints from $\mathrm{BR}(B_s \to \mu^+\mu^-)$ and the double ratio $R$ in the GMSB parameter planes $(\Lambda,\tan\beta)$ on the left and $(M_{mess},\tan\beta)$ on the right. The colour coding is given in section \ref{sec:cmssm}.}
\label{fig:gmsb_mup}
\end{center}
\end{figure}

\begin{figure}[!t]
\begin{center}
\includegraphics[width=8.cm]{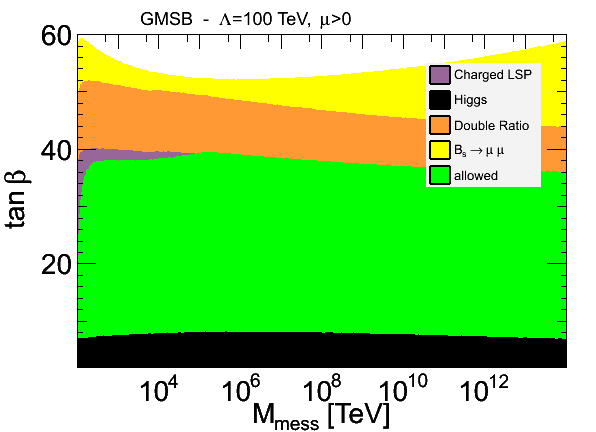}~~~~\includegraphics[width=8.cm]{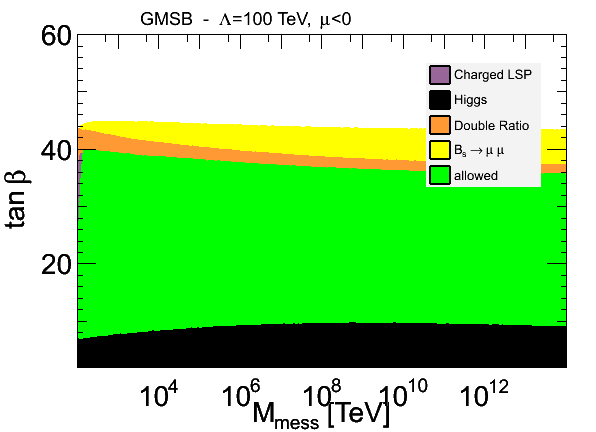}
\caption{Constraints from $\mathrm{BR}(B_s \to \mu^+\mu^-)$ and the double ratio $R$ in the GMSB parameter 
plane $(M_{mess},\tan\beta)$ with $\Lambda=100$ TeV, for $\mu>0$ (on the left) and $\mu<0$ (on the right).}
\label{fig:gmsb_L100}
\end{center}
\end{figure}

In Fig.~\ref{fig:gmsb_mup} we show the results in the parameter planes $(\Lambda,\tan\beta)$ and $(M_{mess},\tan\beta)$. Again, the region of large $\tan\beta$ is 
the most restricted by the flavour observables. 
To see better the regions in the parameter space which are 
excluded by $\mathrm{BR}(B_s \to \mu^+\mu^-)$ and the double ratio $R$, we show in Fig.~\ref{fig:gmsb_L100} the results in the plane $(M_{mess},\tan\beta)$ 
for a fixed value of $\Lambda=100$ TeV for both $\mu>0$ and $\mu<0$. It is remarkable to see that $\tan\beta \gtrsim 40$ is excluded regardless of the value 
of $M_{mess}$, 
while the same plane is probed by the well-known $\mathrm{BR}(B\to X_s\gamma)$ constraints only for a very large messenger scale 
($M_{mess \gtrsim} 10^{10}$ TeV) \cite{Mahmoudi:2007gd}.

\begin{figure}[!t]
\begin{center}
\includegraphics[width=8.cm]{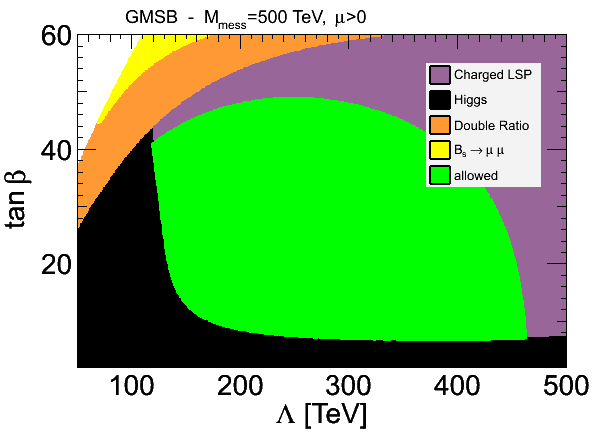}~~~~\includegraphics[width=8.cm]{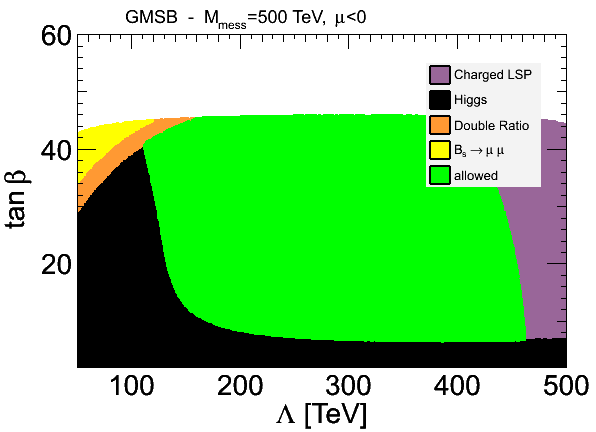}
\caption{Constraints from $\mathrm{BR}(B_s \to \mu^+\mu^-)$ and the double ratio $R$ in the GMSB parameter plane $(\Lambda,\tan\beta)$ with $M_{mess}=500$ TeV, for $\mu>0$ (on the left) and $\mu<0$ (on the right).}
\label{fig:gmsb_M500}
\end{center}
\end{figure}

\newpage
Fig.~\ref{fig:gmsb_M500} presents the constraints in the plane $(\Lambda,\tan\beta)$  with $M_{mess}=500$ TeV and shows that only relatively small values of $\Lambda$ are affected by $B_s \to \mu^+\mu^-$. In the white area (which is especially large in the case of $\mu<0$) it is not possible to find any valid model point.

\newpage
\subsection{CNMSSM}
The last scenario that we consider is a constrained version of the NMSSM (CNMSSM) with semi-universal parameters defined at the GUT scale \cite{NMSSM}. 
The choice of a semi-universal scenario instead of the case of strict universality 
facilitates the obtention of valid NMSSM points \cite{Belanger:2008nt}. 
In this scenario, $\kappa$, $\lambda$ and $m_S^2$ are computed from the minimization equations and the free parameters are $\{m_0, m_{1/2}, A_0, A_\kappa, \lambda, \tan\beta, \mbox{sgn}(\mu)\}$. Previous studies were performed in \cite{Domingo:2007dx,Mahmoudi:2010xp}.
Our sample of 300,000 random points is generated in the ranges $m_0 \in [50,2000]$ GeV, $m_{1/2} \in [50,2000]$ GeV, $A_0 \in [-2000,2000]$ GeV, $A_\kappa \in [-2000,2000]$ GeV, $\lambda \in [10^{-3},1]$ and $\tan\beta \in [1,60]$.

\begin{figure}[!t]
\begin{center}
\includegraphics[width=8.cm]{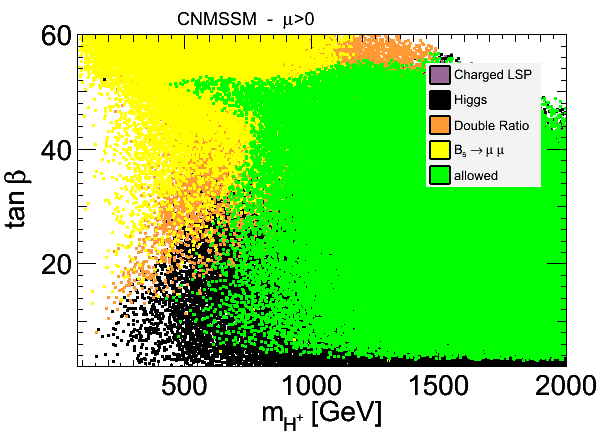}~~~~\includegraphics[width=8.cm]{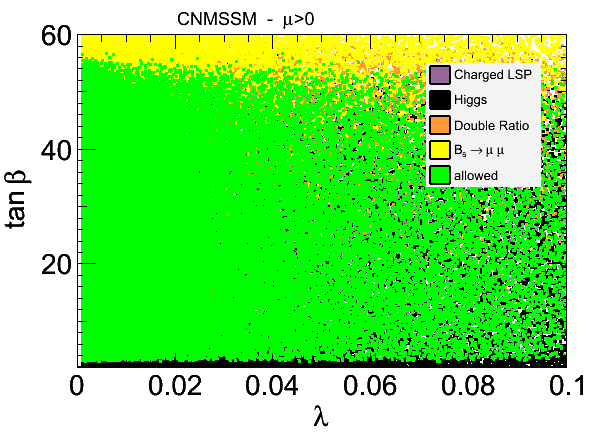}
\caption{Constraints from $\mathrm{BR}(B_s \to \mu^+\mu^-)$ and the double ratio $R$ in CNMSSM in the parameter planes $(m_{H^+},\tan\beta)$ and $(\lambda,\tan\beta)$. The colour coding is given in section \ref{sec:cmssm}.}
\label{fig:cnmssm}
\end{center}
\end{figure}

\begin{figure}[!th]
\begin{center}
\includegraphics[width=8.cm]{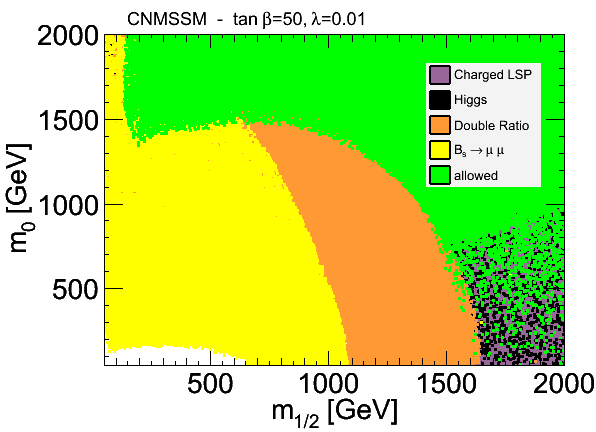}~~~~\includegraphics[width=8.cm]{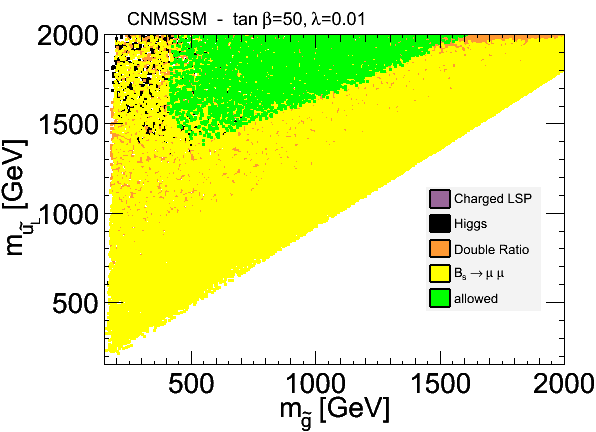}\\[6mm]
\includegraphics[width=8.cm]{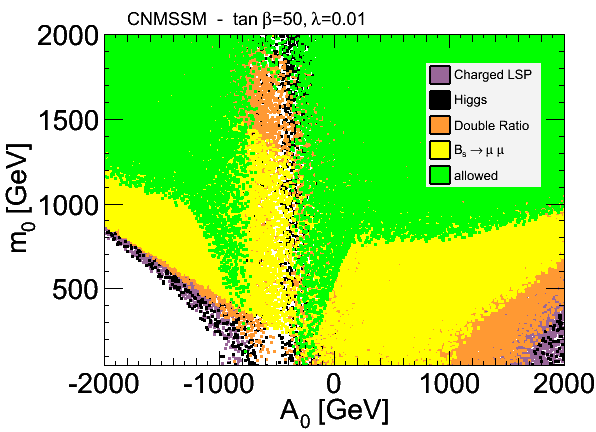}~~~~\includegraphics[width=8.cm]{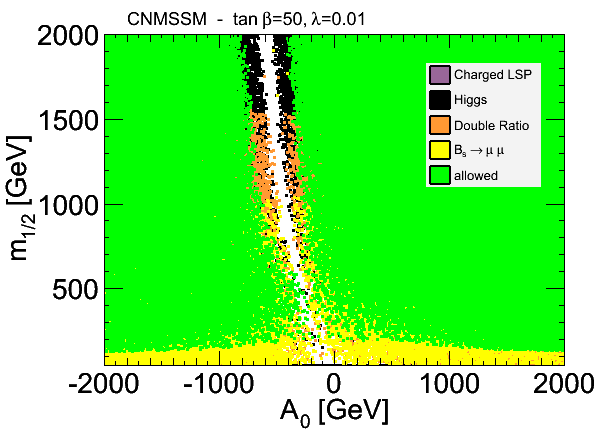}\\[6mm]
\includegraphics[width=8.cm]{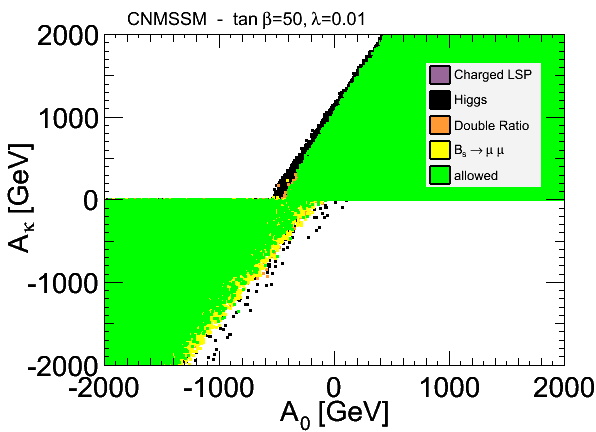}~~~~\includegraphics[width=8.cm]{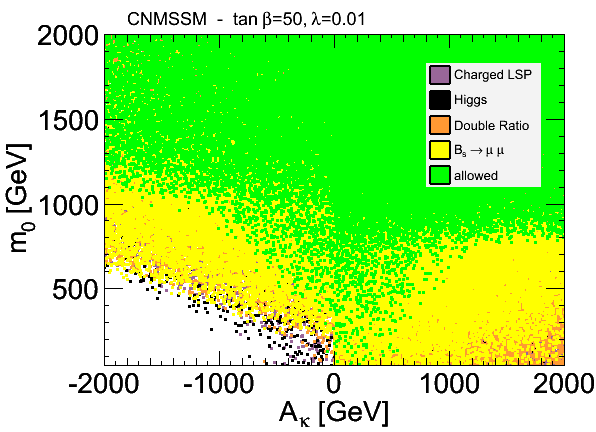}
\caption{Constraints from $\mathrm{BR}(B_s \to \mu^+\mu^-)$ and the double ratio $R$ in different CNMSSM parameter planes for $\lambda=0.01$ and $\tan\beta=50$ with $\mu>0$.}
\label{fig:cnmssm_l01}
\end{center}
\end{figure}

\begin{figure}[!th]
\begin{center}
\includegraphics[width=8.cm]{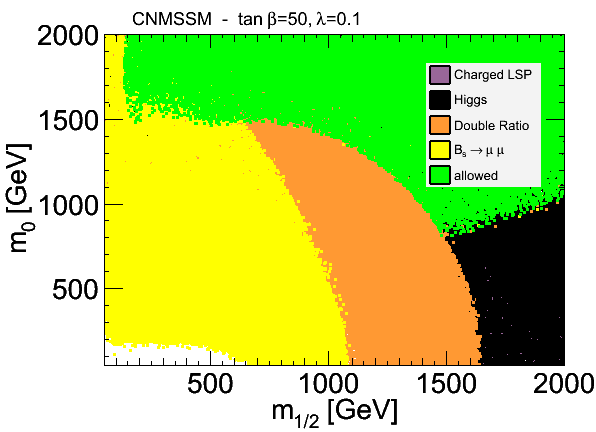}~~~~\includegraphics[width=8.cm]{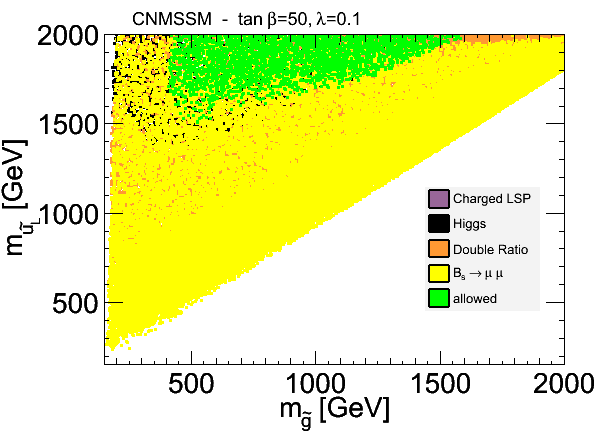}\\[6mm]
\includegraphics[width=8.cm]{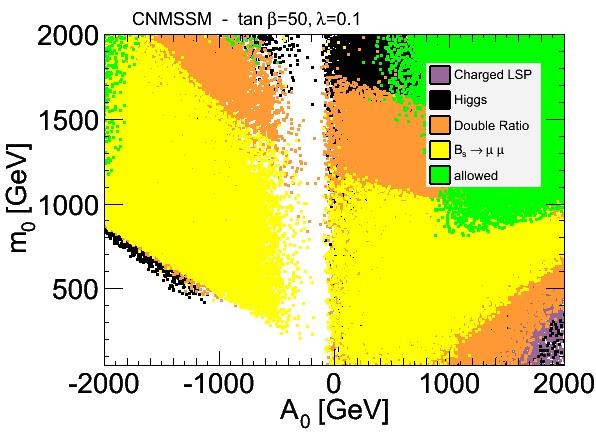}~~~~\includegraphics[width=8.cm]{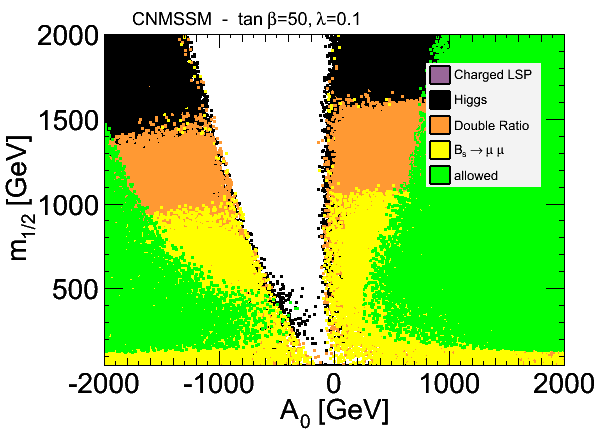}\\[6mm]
\includegraphics[width=8.cm]{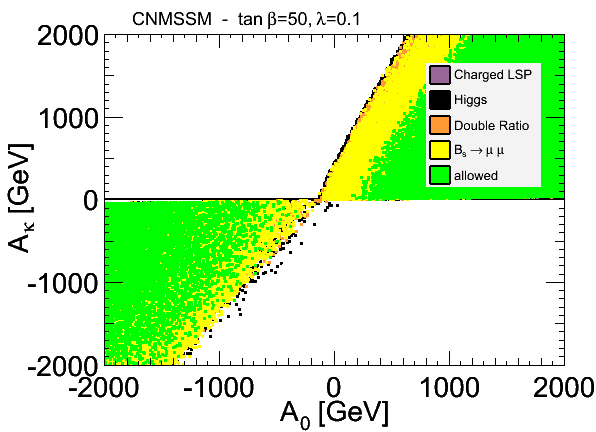}~~~~\includegraphics[width=8.cm]{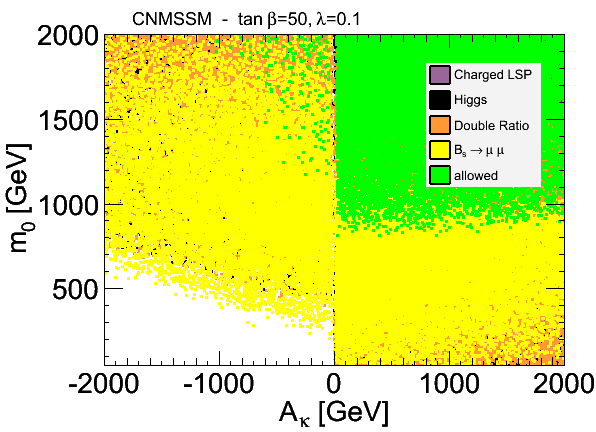}
\caption{Constraints from $\mathrm{BR}(B_s \to \mu^+\mu^-)$ and the double ratio $R$ in different CNMSSM parameter planes for $\lambda=0.1$ and $\tan\beta=50$ with $\mu>0$.}
\label{fig:cnmssm_l1}
\end{center}
\end{figure}

The results are displayed in Fig.~\ref{fig:cnmssm} in the parameter planes $(m_{H^+},\tan\beta)$ and $(\lambda,\tan\beta)$. The constraints are more severe for large $\tan\beta$, small $m_{H^+}$ and large $\lambda$.
In Fig.~\ref{fig:cnmssm_l01} we fix two of the parameters, namely $\lambda=0.01$ and $\tan\beta=50$. This allows us to see in a clearer way the effect of the constraints on the other parameters. In Fig.~\ref{fig:cnmssm_l1} the same results are shown for $\lambda=0.1$. As mentioned before, the constraints are more pronounced for larger $\lambda$.

\begin{figure}[!th]
\begin{center}
\includegraphics[width=8.cm]{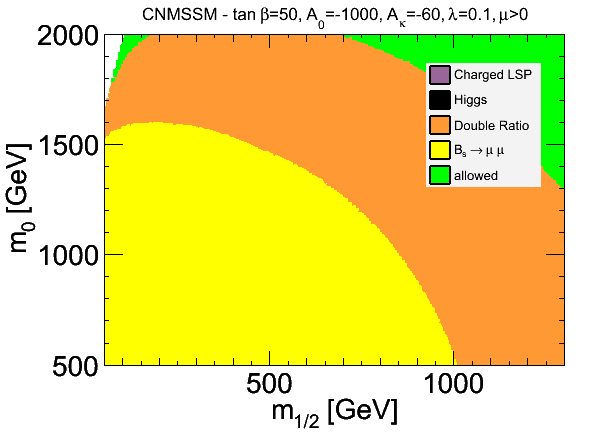}
\caption{Constraints from $\mathrm{BR}(B_s \to \mu^+\mu^-)$ and the double ratio $R$ in the CNMSSM parameter plane $(m_{1/2},m_0)$ for $A_0=1000$ GeV, $A_\kappa=-60$ GeV, $\tan\beta=50$ and $\lambda=0.1$.}
\label{fig:cnmssm_tb50}
\end{center}
\end{figure}

As a final example we fix all the parameters except two, to see the results in a two-dimensional plane. This is done in Fig.~\ref{fig:cnmssm_tb50} for $A_0=1000$ GeV, $A_\kappa=-60$ GeV, $\tan\beta=50$ and $\lambda=0.1$. As can be seen, a large part of this parameter plane is excluded by $\mathrm{BR}(B_s \to \mu^+\mu^-)$ and the double ratio $R$.

\newpage
\subsection{Discussion}

\begin{figure}[!t]
\begin{center}
\includegraphics[width=8.cm]{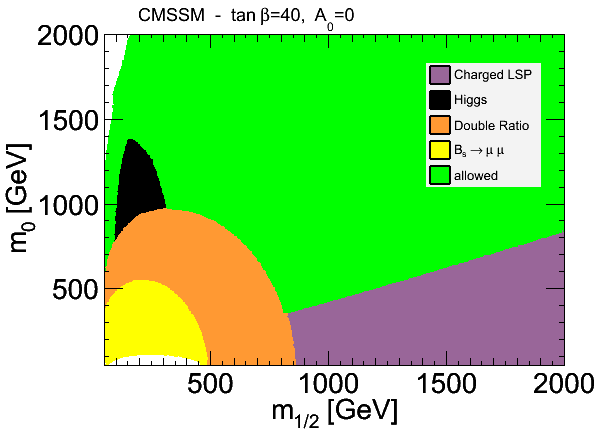}~~~~\includegraphics[width=8.cm]{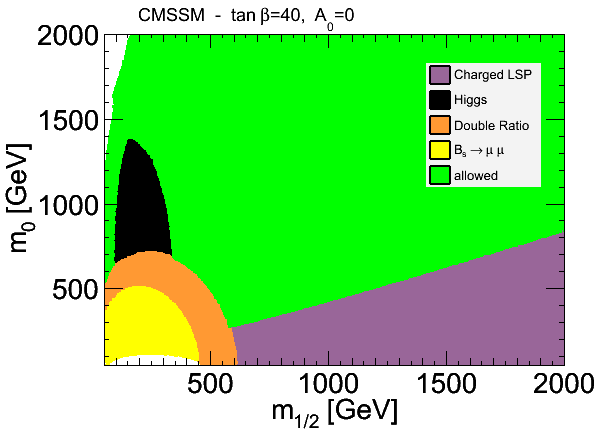}
\caption{Constraints from $\mathrm{BR}(B_s \to \mu^+\mu^-)$ and the double ratio $R$ in the CMSSM parameter planes $(m_{1/2},m_0)$ for $\tan\beta=40$.
On the left, the most constraining case with low $|V_{ub}|$ and high $f_{B_s}$ and on the right the least constraining case with high $|V_{ub}|$ and low $f_{B_s}$.}
\label{fig:cmssm_bw}
\end{center}
\end{figure}
\begin{figure}[!t]
\begin{center}
\includegraphics[width=8.cm]{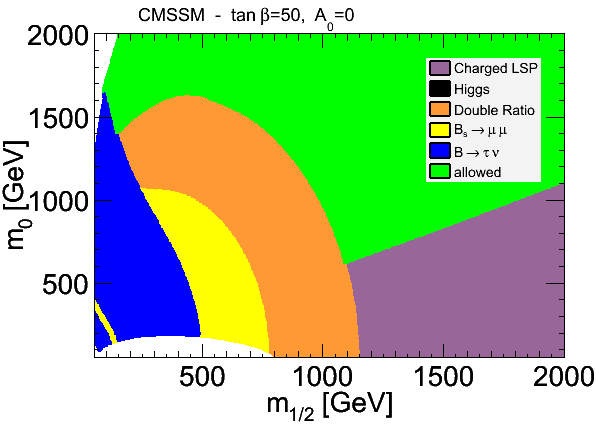}
\caption{Constraints from $\mathrm{BR}(B_s \to \mu^+\mu^-)$, the double ratio $R$, and  $B_u\to \tau\nu$ 
in the CMSSM plane $(m_{1/2},m_0)$, for  $\tan\beta=50$ and $A_0=0$ GeV. This figure is the same as Fig.~\ref{fig:cmssm_mup} 
(upper left panel) but with
the constraint from $B_u\to \tau\nu$ superimposed.}
\label{fig:btaunu}
\end{center}
\end{figure}

In the above subsections, we investigated the constraining power of $\mathrm{BR}(B_s \to \mu^+\mu^-)$ and the double ratio $R$ for different SUSY scenarios. As explained in sections \ref{sec:Bsmumu} and \ref{sec:doubleratio}, the main input parameter for $\mathrm{BR}(B_s \to \mu^+\mu^-)$ is $f_{B_s}$, while $|V_{ub}|$ is the most important input for $R$. To examine how the choice of these input parameters can affect our results, we consider here two scenarios, namely the ``least constraining'' (with high $|V_{ub}|$ and low $f_{B_s}$) and ``most constraining'' (with low $|V_{ub}|$ and high $f_{B_s}$) cases for $\mathrm{BR}(B_s \to \mu^+\mu^-)$ and the double ratio. For the least constraining scenario we consider the inclusive determination of $|V_{ub}|$ with the central value being $4.34 \times 10^{-3}$ \cite{Asner:2010qj}, and $f_{B_s}=232$ MeV \cite{:2011gx}. For the most constraining case we take the exclusive value $|V_{ub}|=3.42 \times 10^{-3}$ \cite{Asner:2010qj} and $f_{B_s}=250$ MeV \cite{Simone:2010zz}. To compare these two cases we take an example in the CMSSM scenario 
with $\tan\beta=40$ and $A_0=0$. The results are presented in Fig.~\ref{fig:cmssm_bw}. As can be seen, in the most constraining case, the exclusion limits are greatly increased while in the least constraining case the results are only slightly changed. This shows that the analysis in the previous subsections does not correspond to a particularly optimistic choice of the input parameters.

The next point we discuss here is the effect of $B_u\to \tau\nu$ in the double ratio. The constraints from $B_u\to \tau\nu$ alone on the parameter space of $[m_0,m_{1/2}]$ have been presented in several works
(e.g. \cite{Mahmoudi:2008tp}) and the excluded region differs from that obtained from $\mathrm{BR}(B_s\to\mu^+\mu^-)$ alone, as can be seen in Fig.~\ref{fig:btaunu}. 
In most of the parameter space, BR($B_u\to \tau\nu$) is reduced with respect to the SM value, leading to the large blue excluded strip in Fig.~\ref{fig:btaunu}. On the other hand, in the small strip, BR($B_u\to \tau\nu$) is larger than in the SM. In the narrow region in between, a cancellation happens since the charged Higgs contribution is roughly twice that of the SM contribution and so $B_u\to \tau\nu$ cannot exclude this parameter space.
As can be seen from the figure, $\mathrm{BR}(B_s\to\mu^+\mu^-)$ probes larger values of $m_{1/2}$ than $B_u\to \tau\nu$, although $B_u\to \tau\nu$ can exclude part of the region $1300 \,{\rm GeV} < m_0 < 1600 \,{\rm GeV}$ and $m_{1/2}< 200$ GeV which cannot be excluded from $\mathrm{BR}(B_s\to\mu^+\mu^-)$ alone and the double ratio.

The reason why the double ratio is more constraining than $\mathrm{BR}(B_s\to\mu^+\mu^-)$ alone is mainly due to two reasons:
i) $|V_{ub}|$ is used as an input parameter in the double ratio, instead of $f_{B_s}$. Although these two parameters have comparable errors, their current central values give rise to stronger constraints from 
the double ratio, as discussed in the preceding paragraph.
This could not have been expected, and a value of $f_{B_s}$ much larger than that preferred by
lattice QCD would have ensured that $\mathrm{BR}(B_s\to\mu^+\mu^-)$ alone had the stronger constraints;
ii) The experimental value of BR($B_u\to \tau\nu$), which enters the derivation of $\eta$ in Eq.~(\ref{ratio_eta}), is larger than the
SM expectation, and so reduces $R$ in Eq.~(\ref{R_doubrat}), leading to a stronger constraint on the SUSY parameter space.
The SUSY contributions to $B_u\to \tau\nu$ gives rise to a scale factor which multiplies BR($B_u\to \tau\nu$). 
When we manually set this scale factor to be equal to 1 (as in the SM), the excluded region of the
plane  $[m_0,m_{1/2}]$ does not change much. Therefore we conclude that the points i) and ii) above are the main reasons why the double ratio gives the superior constraints.

\begin{figure}[!t]
\begin{center}
\includegraphics[width=8.cm]{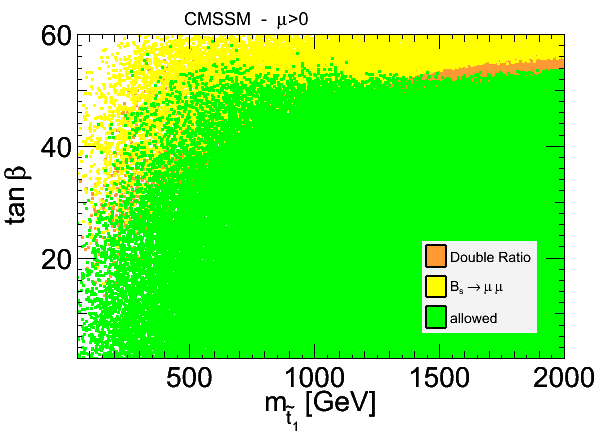}~~~~\includegraphics[width=8.cm]{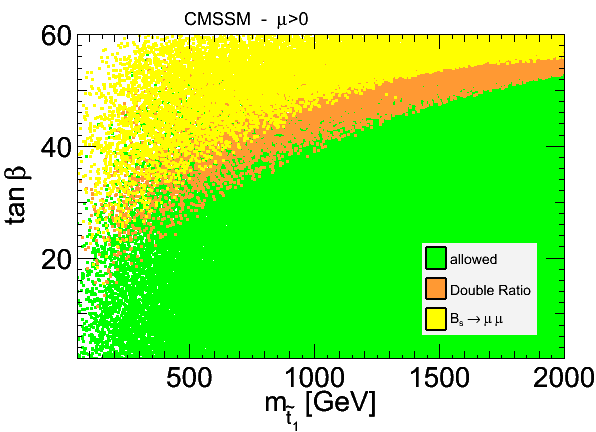}
\includegraphics[width=8.cm]{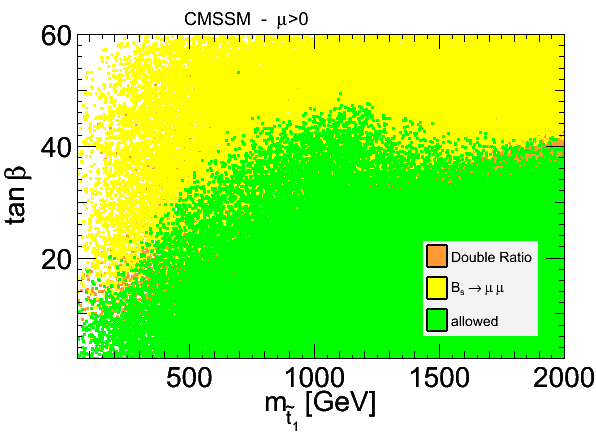}~~~~\includegraphics[width=8.cm]{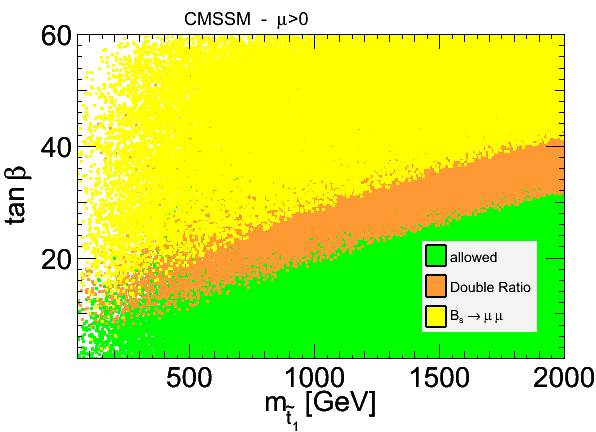}
\caption{Constraints from $\mathrm{BR}(B_s \to \mu^+\mu^-)$ and the double ratio $R$ in the CMSSM parameter planes $(m_{\tilde{t}_1},\tan\beta)$ in the hypothetical case of a SM-like measurement (lower panel) and with the current experimental limits (upper panel). In the left panel 
the allowed points in green are displayed in the background while in the right panel they are in foreground.}
\label{fig:cmssm_sm}
\end{center}
\end{figure}

Finally we discuss the effect of a hypothetical measurement of $\mathrm{BR}(B_s \to \mu^+\mu^-)$ at the SM value $(3.5\pm0.3)\times 10^{-9}$. Fig.~\ref{fig:cmssm_sm} shows the obtained impact in the CMSSM plane $(m_{\tilde{t}_1},\tan\beta)$ with all the parameters being varied in the intervals given in section~\ref{sec:cmssm}. For comparison, the same parameter plane with the current experimental limits is also provided. As can be seen, almost no scenario with $\tan\beta\gtrsim 45$ remains viable regardless of the other parameters in the case of a SM-like discovery, and the parameter space of the CMSSM becomes very restricted. 

\section{Experimental prospects}
\label{sec:prospects}

At present, the best upper limit for $\mathrm{BR}(B_s\to\mu^+\mu^-)$ measured in a single experiment comes from LHCb \cite{LHCb_alone}:
\begin{equation}
\mathrm{BR}(B_s\to\mu^+\mu^-) < 1.5 \times 10^{-8}
\end{equation}
at 95\% C.L. This upper limit is followed closely by the result from CMS \cite{Chatrchyan:2011kr}:
\begin{equation}
\mathrm{BR}(B_s\to\mu^+\mu^-) < 1.9 \times 10^{-8}
\end{equation}
at 95\% C.L. These two results were officially combined for EPS conference
in \bibref{CMS_plus_LHCb}, giving the upper limit of
\begin{equation}
\mathrm{BR}(B_s\to\mu^+\mu^-) < 1.1 \times 10^{-8}\;,
\label{combined_limit}
\end{equation}
which we will use to constrain the parameter space of SUSY models. The CDF collaboration obtains a 95\% C.L. upper limit \cite{Aaltonen:2011fi}: 
\begin{equation}
\mathrm{BR}(B_s\to\mu^+\mu^-) < 4.0 \times 10^{-8}\;,
\end{equation}
together with a one sigma interval 
\begin{equation}
\mathrm{BR}(B_s\to\mu^+\mu^-)  = (1.8^{+1.1}_{-0.9})\times10^{-8}\;,
\end{equation}
coming from an observed excess over the expected background which corresponds to a $p-$value of 0.27$\%$.
Finally, the D0 collaboration obtains the 95\% C.L. upper limit \cite{Abazov:2010fs}: 
\begin{equation}
\mathrm{BR}(B_s\to\mu^+\mu^-) < 5.1 \times 10^{-8}\;.
\end{equation}

The preliminary result on $\mathrm{BR}(B_s \to \mu^+\mu^-)$ \cite{CMS_plus_LHCb}
from the combination of the limits from LHCb and CMS shows an excess of more than one sigma (\CLb $\approx 0.92$ for values of the BR around
the SM value) with respect to the background-only hypothesis. This excess can be accounted for by a 
$\mathrm{BR}(B_s \to \mu^+\mu^-) \approx (3.7^{+3.7}_{-2.7})\times 10^{-9}$.
However, the signal significance is not enough to claim evidence. 
In this section we study the experimental sensitivity to \Bsmumu
and the prospects for its measurement in the period of operation of the LHC at $\sqrt s=7$ TeV.

\subsection{Combination LHC-CDF}
\label{sec:avec_CDF}
The CDF experiment at the Tevatron has reported a $p-$value of 0.27$\%$ for the background only hypothesis \cite{Aaltonen:2011fi}. 
In order 
to evaluate whether a combination of results from CMS, LHCb and CDF could lead to evidence for a signal, we perform an approximate
combination of the results of the three experiments, based on the signal and background expectations and the observed
pattern of events. We use \texttt{mc\_limit} \cite{mc_limit} to combine the results of the different experiments and to
extract the confidence levels. We have also scaled $f_d/f_s$ to the value measured at LHCb \cite{fdfs} in order to be consistent
with the value used in the LHC combined result.

According to this study, a hypothetical combination of the LHCb and CMS results with that of CDF would increase  \CLb to $\sim 0.994$
(for values of the BR close to the most probable value), 
which is close to a 3$\sigma$ deviation. Note that this is approximately the same signal significance that CDF obtains alone.
This approximate study leads to the following averaged branching ratio:
\begin{equation}
\mathrm{BR}(B_s \to \mu^+\mu^-)_{CDF+LHC} \approx (6^{+5}_{-3}) \times 10^{-9}\;.
\label{eq:CDF_LHCb_CMS}
\end{equation}

However, at the time of writing this paper, this kind of combination has not been performed officially.

\subsection{Sensitivity to BR(\Bsmumu) at the LHC}
\label{sec:discovery1}
\begin{figure}[!t]
\begin{center}
\includegraphics[width=8.cm]{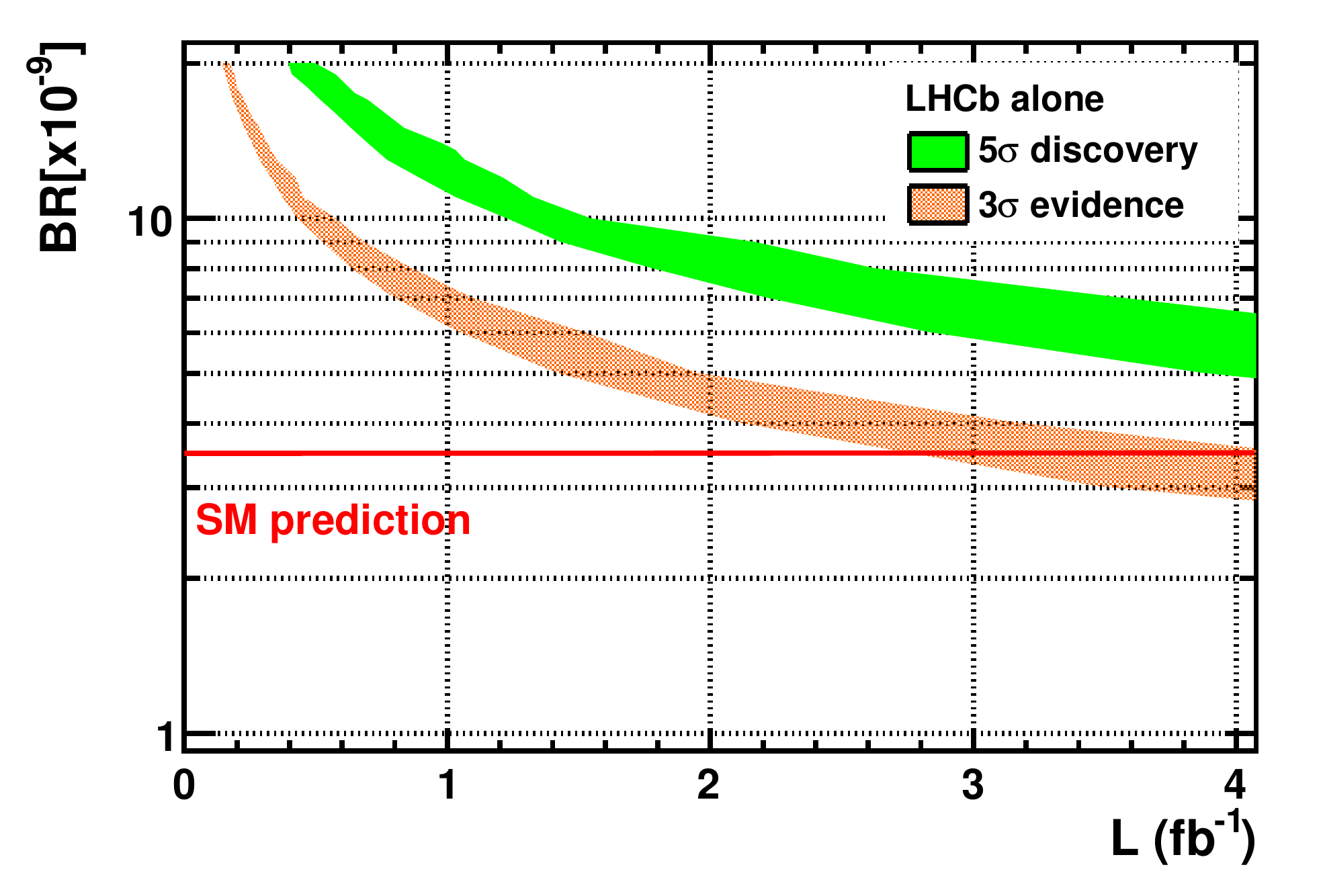}~~~~\includegraphics[width=8.cm]{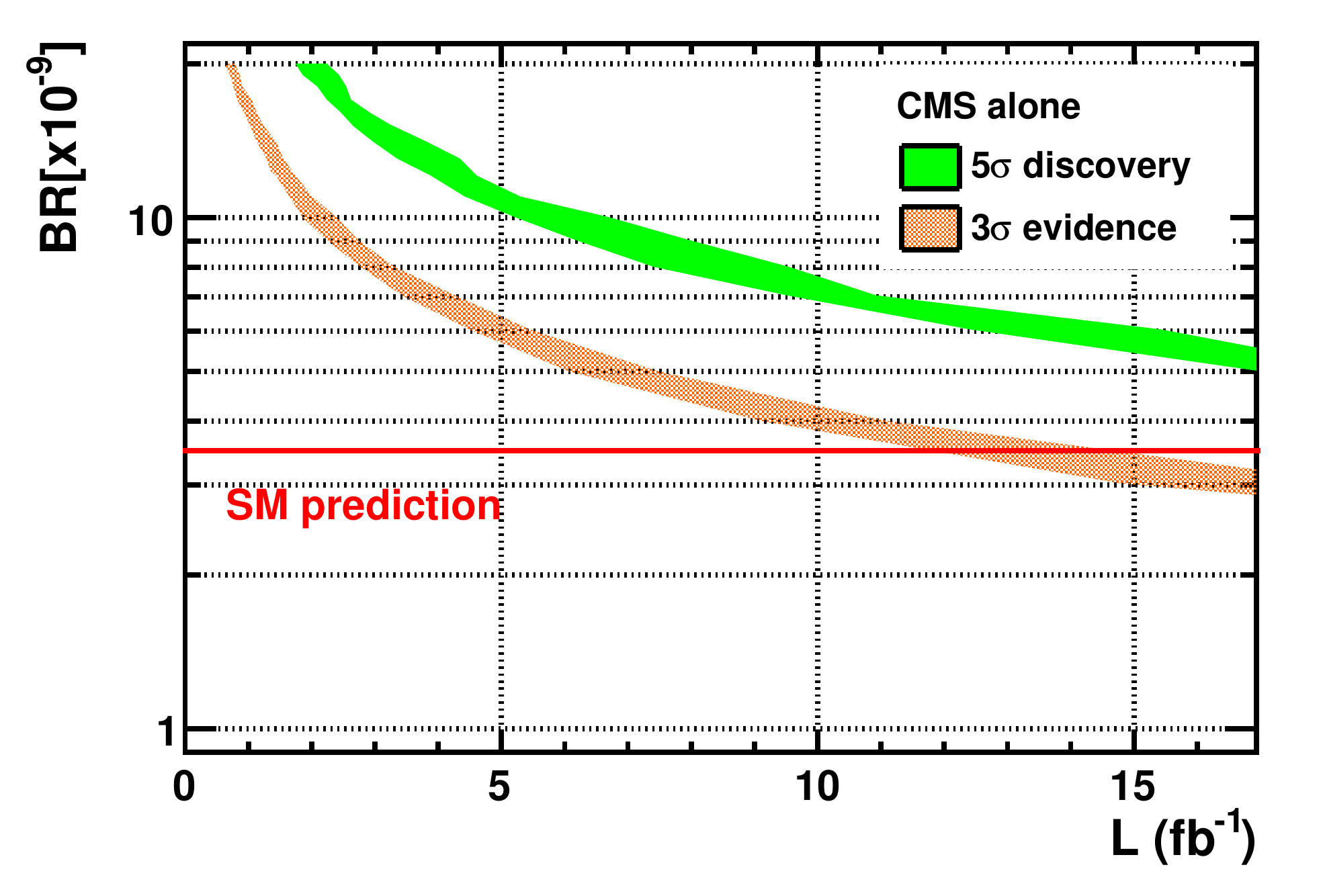}
\caption{Required luminosity in order to provide a 3$\sigma$ evidence (orange) or a 5$\sigma$ discovery (green)
of a given $\mathrm{BR}(B_s \to \mu^+\mu^-)$ on the left for LHCb and on the right for CMS.}
\label{fig:sensitivity_alone}
\end{center}
\end{figure}

We perform a toy MC study in order to determine how much luminosity is needed to obtain evidence for $B_s \to \mu^+\mu^-$
at the LHC. For this, we scale the signal and background expectations accordingly with the increase of luminosity.
\figref{fig:sensitivity_alone} shows the integrated luminosity that is needed in order to obtain a 3(5) $\sigma$ evidence
(discovery) of a given $\mathrm{BR}(B_s \to \mu^+\mu^-)$ in either LHCb or CMS.

\begin{figure}[!t]
\begin{center}
\includegraphics[width=8.cm]{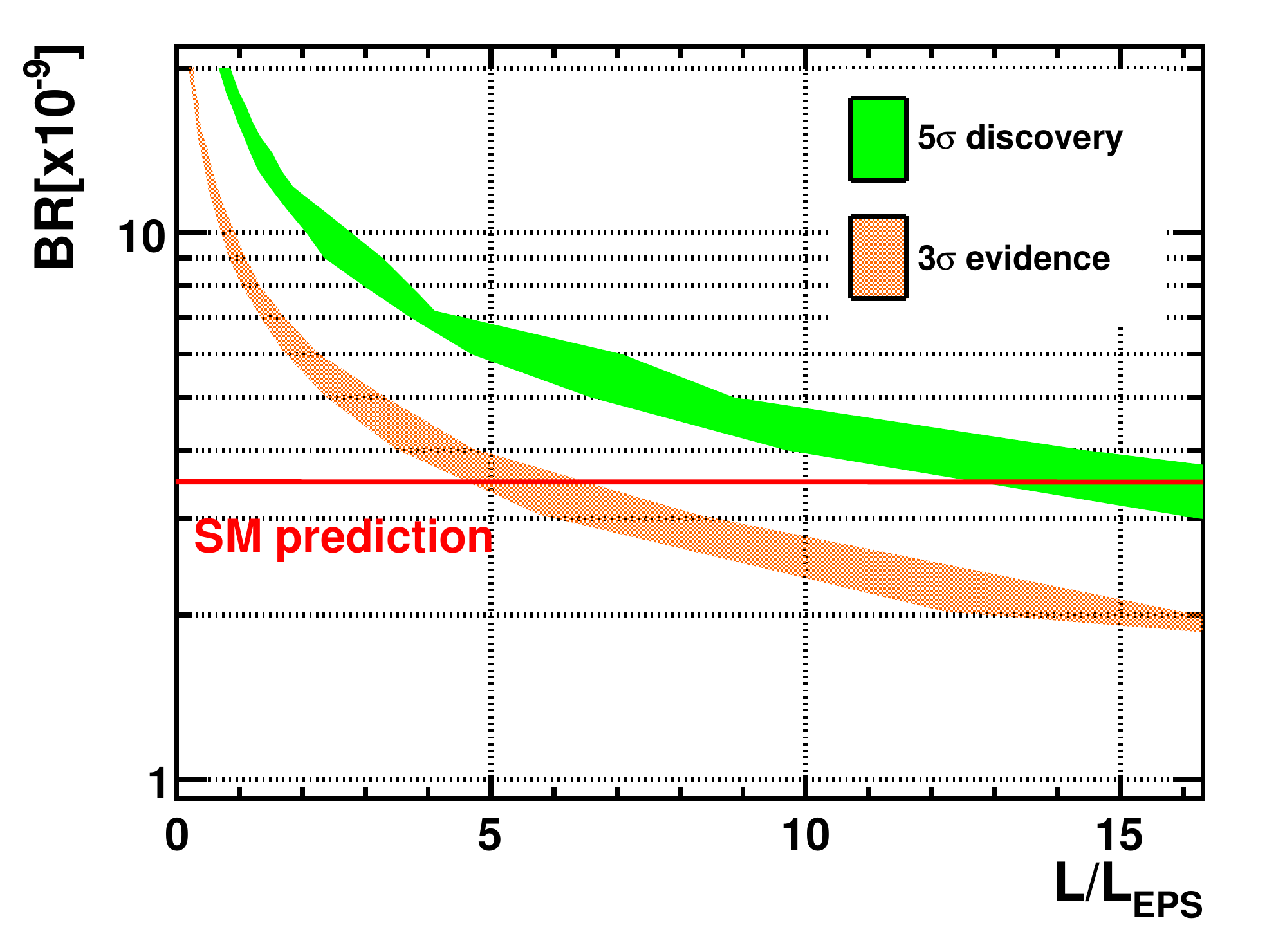}
\caption{Required luminosity in order to provide a 3$\sigma$ evidence (orange) or a 5$\sigma$ discovery (green)
of a given $\mathrm{BR}(B_s \to \mu^+\mu^-)$ for LHCb and CMS combined. The luminosity is expressed in terms of the luminosity used 
in \cite{CMS_plus_LHCb}, (0.34 \invfb for LHCb
and 1.14 \invfb for CMS).}
\label{fig:expected_sensitivity}
\end{center}
\end{figure}

Assuming that the ratio of luminosities between CMS 
and LHCb remains at the value of the current analysis 
({\it i.e.} CMS takes approximately four times more data than LHCb over the same period of time), we show in 
\figref{fig:expected_sensitivity} the integrated luminosity scale factor (with respect to the amount of data used in
\cite{CMS_plus_LHCb}) that would be needed
for the discovery of a given $\mathrm{BR}(B_s \to \mu^+\mu^-)$ in the case of a CMS$+$LHCb combination.
The width of the bands reflects possible scenarios for the evolution of the systematic uncertainties,
where the lower side assumes negligible systematics and the upper side assumes that the systematics do not get reduced with time.
It can be seen that with 6-8 times more luminosity than that used in \bibref{CMS_plus_LHCb} a CMS$+$LHCb combination could provide 
evidence at the 3$\sigma$ level for $\mathrm{BR}(B_s \to \mu^+\mu^-)$ of the SM.
This corresponds to between 2 and 3 \invfb for LHCb and between 7 and 10 \invfb for CMS. As the sensitivity
of CMS is equivalent to that of LHCb for four times more luminosity, a scenario in which CMS takes up to 14 \invfb and LHCb
takes 2 \invfb would afford equal sensitivity as a combination of CMS with 10 \invfb and LHCb with 3 \invfb.
From this toy MC study we conclude that the SM prediction for $\mathrm{BR}(B_s \to \mu^+\mu^-)$ 
is likely to be probed during the operation of the LHC at $\sqrt s=7$ TeV (i.e. before the end of the year 2012). 
If ATLAS can manage to obtain sensitivity to $\mathrm{BR}(B_s \to \mu^+\mu^-)$ which is comparable to that
of CMS, then even a 5$\sigma$ discovery for 
a SM-like $\mathrm{BR}(B_s \to \mu^+\mu^-)$ would be possible during the run at $\sqrt s=7$ TeV.
However, from pre-LHC MC studies in \bibref{thesis_2010_068} the sensitivity of ATLAS was found to be inferior to that of CMS.
If experimental evidence of \Bsmumu is achieved at the LHC, the double ratio in Eq.~(\ref{doub-rat}) would be
measured for the first time. Moreover, 
limits on the ratio $\mathrm{BR}(B_d \to \mu^+\mu^-)/\mathrm{BR}(B_s \to \mu^+\mu^-)$  
(which is a very
interesting test of Minimal Flavour Violation) would also be set. 
If $\mathrm{BR}(B_s \to \mu^+\mu^-)$ is much smaller than the SM prediction 
(as can happen for example in the MSSM \cite{Dedes:2008iw} and NMSSM), values down to $O(5\times10^{-10})$
can still be discovered with an upgrade of the LHCb.

\subsection{NP discovery with \Bsmumu}
\begin{figure}[!th]
\begin{center}
\includegraphics[width=8.cm]{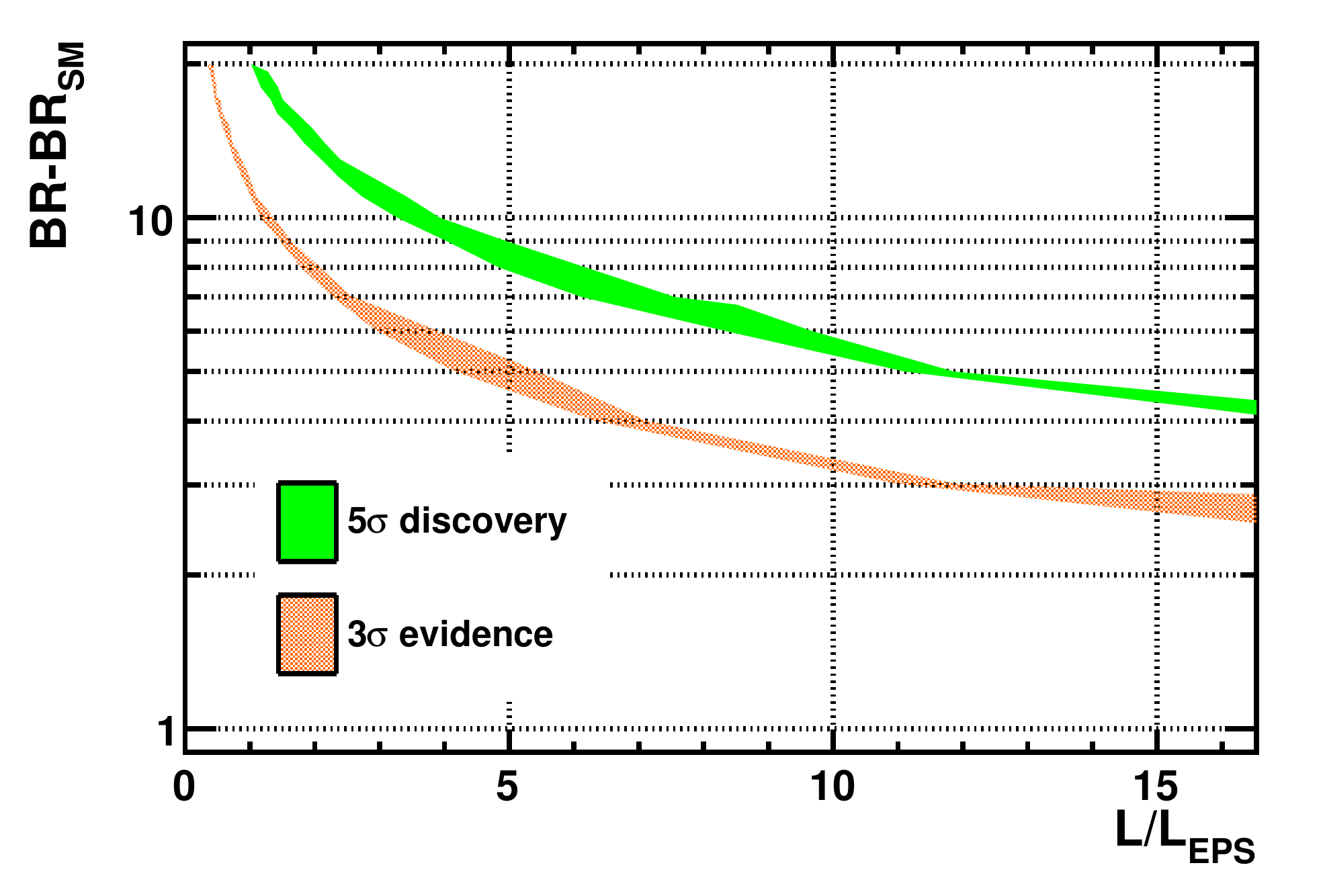}
\caption{Required luminosity in order to provide a 3$\sigma$ evidence (orange) or a 5$\sigma$ discovery (green)
of a given NP contribution to $\mathrm{BR}(B_s \to \mu^+\mu^-)$ for LHCb and CMS combined. The luminosity is expressed in terms of the luminosity used for \bibref{CMS_plus_LHCb}, (0.34 \invfb for LHCb
and 1.14 \invfb for CMS).}
\label{fig:NP_discovery}
\end{center}
\end{figure}

In \secref{sec:discovery1}  we discussed the luminosity needed for discovery of \Bsmumu. However, a measurement of \Bsmumu
with a branching ratio larger than the SM prediction does not necessarily mean a New Physics (NP) discovery. In such a case, the compatibility with the 
SM prediction has to be computed. \figref{fig:NP_discovery} is the equivalent of \figref{fig:expected_sensitivity} but with the SM
rate for $\mathrm{BR}(B_s \to \mu^+\mu^-)$ being considered
as a background, and the signal corresponds to the NP contribution to $\mathrm{BR}(B_s \to \mu^+\mu^-)$.  We can see that for the same luminosity 
needed for a 3$\sigma$ evidence of a SM-like signal, the LHC could alternatively claim NP at $3\sigma$ if the NP contribution is
of the order of $4-5\times10^{-9}$, {\it i.e}, if the actual $\mathrm{BR}(B_s \to \mu^+\mu^-)$ is O(8$\times10^{-9}$). 
Finally, with the current uncertainties in $f_d/f_s$  ($7.9\%$) and in the SM prediction ($8\%$), only values of $\mathrm{BR}(B_s \to \mu^+\mu^-)$ that are at least
33\%(55)$\%$ larger than the SM prediction can allow exclusion of a SM-like rate at 3(5)$\sigma$.

\section{Conclusions}

The decay $B_s\to \mu^+\mu^-$ is known to be a very effective probe of SUSY models with large ($> 30$) $\tan\beta$, and 
its importance has been emphasised in numerous studies over the past decade.
Due to its distinct signature, this decay can be searched for by three LHC collaborations: LHCb, CMS and ATLAS.
Recently, searches by LHCb and CMS have been released, and have improved the upper limit on
its branching ratio to BR($B_s\to \mu^+\mu^-)< 1.1\times 10^{-8}$. Using this new bound, we performed a study of the constraints
on the parameter space of five distinct SUSY models. We emphasised that such indirect constraints can be stronger
than those which are obtained from the ongoing direct searches for SUSY particles at the LHC.
For instance, in the CMSSM for $\tan\beta \sim 50$, the SUSY particles have to be very heavy and in particular squarks cannot be lighter than $\sim 1.2 -2$ TeV in order to be compatible with the upper limit on BR($B_s\to \mu^+\mu^-)$. Nevertheless, in the scenarios we investigated here, in spite of the severe constraints we obtained, there is still room for SUSY contributions in large parts of the parameter space, especially for small $\tan\beta$.

In addition, we considered an alternative observable which includes BR($B_s\to \mu^+\mu^-)$, 
namely a double ratio formed from the decays $B_s\to \mu^+\mu^-, B_u\to \tau\nu, D\to \mu\nu$ and $D_s\to \mu\nu/\tau\nu$.
The magnitude of the double ratio depends on the CKM matrix element $|V_{ub}|$, a parameter for which there is already considerable experimental information,
and the prospects for further precision in measurements of $|V_{ub}|$ are promising. In contrast, the magnitude of BR($B_s\to \mu^+\mu^-)$ 
depends on the absolute value of the decay constant $f_{B_s}$, and thus a comparative study of the constraints obtained from 
these two observables is instructive. We showed that the double ratio can provide stronger constraints on the SUSY parameter space, and 
we advocate its use when discussing the impact of BR($B_s\to \mu^+\mu^-)$ alone on SUSY models.

The final integrated luminosity of the operation of the LHC at $\sqrt s =7$ TeV is likely to be
significantly larger than the amount that was anticipated at the start of the run. 
Both CMS and LHCb will have a chance to obtain a significant signal 
by the end of the run, even if BR($B_s\to \mu^+\mu^-$) is as small as the prediction in the SM. Throughout the  
run at $\sqrt s =7$ TeV, the ongoing searches for $B_s\to \mu^+\mu^-$ will continue to 
compete with the direct searches for SUSY particles as a probe of the parameter space of SUSY models.

\section*{Acknowledgements}
A.G.A was supported by a Marie Curie 
Incoming International Fellowship, FP7-PEOPLE-2009-IIF, Contract No. 252263.



\begin{thebibliography}{999}

\bibitem{Carena:2006ai}
  M.~S.~Carena, A.~Menon, R.~Noriega-Papaqui, A.~Szynkman and C.~E.~M.~Wagner,
  ``Constraints on $B$ and Higgs physics in minimal low energy supersymmetric
  models'',
  Phys.\ Rev.\  D {\bf 74} (2006) 015009
  [hep-ph/0603106].

\bibitem{Ellis:2007fu}
  J.~R.~Ellis, S.~Heinemeyer, K.~A.~Olive, A.~M.~Weber and G.~Weiglein,
  ``The Supersymmetric Parameter Space in Light of $B^-$ physics Observables
  and Electroweak Precision Data'',
  JHEP {\bf 0708} (2007) 083
  [arXiv:0706.0652].

\bibitem{Mahmoudi:2007gd}
  F.~Mahmoudi,
  ``New constraints on supersymmetric models from $b \to s \gamma$'',
  JHEP {\bf 0712} (2007) 026
  [arXiv:0710.3791].

\bibitem{Heinemeyer:2008fb}
  S.~Heinemeyer, X.~Miao, S.~Su and G.~Weiglein,
  ``$B^-$ Physics Observables and Electroweak Precision Data in the CMSSM,
  mGMSB and mAMSB'',
  JHEP {\bf 0808} (2008) 087
  [arXiv:0805.2359].

\bibitem{Eriksson:2008cx}
  D.~Eriksson, F.~Mahmoudi and O.~St\aa l,
  ``Charged Higgs bosons in Minimal Supersymmetry: Updated constraints and experimental prospects'',
  JHEP {\bf 0811}, 035 (2008)
  [arXiv:0808.3551].

\bibitem{Alok:2009wk}
  A.~K.~Alok and S.~K.~Gupta,
  ``$B_s \to \mu^{+}\mu^{-}$ decay in the R-parity violating minimal supergravity'',
  Eur.\ Phys.\ J.\  C {\bf 65} (2010) 491
  [arXiv:0904.1878].

\bibitem{Choudhury:1998ze}
  S.~R.~Choudhury and N.~Gaur,
  ``Dileptonic decay of $B_s$ meson in SUSY models with large $\tan\beta$'',
  Phys.\ Lett.\  B {\bf 451} (1999) 86
  [hep-ph/9810307].

\bibitem{Babu:1999hn}
  K.~S.~Babu and C.~F.~Kolda,
  ``Higgs mediated $B^0 \to \mu^{+} \mu^{-}$ in minimal supersymmetry'',
  Phys.\ Rev.\ Lett.\  {\bf 84} (2000) 228
  [hep-ph/9909476].

\bibitem{Huang:2000sm}
  C.~-S.~Huang, W.~Liao, Q.~-S.~Yan, S.~-H.~Zhu,
  Phys.\ Rev.\  {\bf D63}, 114021 (2001)
  [hep-ph/0006250].

\bibitem{:2007kv}
  T.~Aaltonen {\it et al.}  [CDF Collaboration],
  ``Search for $B^0_{s} \to \mu^{+} \mu^{-}$ and $B^0_{d} \to \mu^{+} \mu^{-}$
  decays with 2 fb$^{-1}$ of $p \bar{p}$ collisions'',
  Phys.\ Rev.\ Lett.\  {\bf 100} (2008) 101802
  [arXiv:0712.1708].

\bibitem{Abazov:2010fs}
  V.~M.~Abazov {\it et al.}  [D0 Collaboration],
  ``Search for the rare decay $B_s^0 \to \mu^+\mu^-$'',
  Phys.\ Lett.\  B {\bf 693}, 539 (2010)
  [arXiv:1006.3469].

\bibitem{Aaltonen:2011fi}
  T.~Aaltonen {\it et al.}  [CDF Collaboration],
  ``Search for $B_s \to \mu^+\mu^-$ and $B_d \to \mu^+\mu^-$ Decays with CDF
  II'',
  arXiv:1107.2304 [hep-ex].

\bibitem{LHCb_alone}
 R.~Aaij {\it et al.} [LHCb Collaboration], ``Search for the rare decays $B_{s,d} \to \mu^+\mu^-$ with 300 $pb^{-1}$ at LHCb'', LHCb-CONF-2011-037, 27 July 2011.

\bibitem{Chatrchyan:2011kr}
  S.~Chatrchyan {\it et al.}  [CMS Collaboration],
  ``Search for B(s) and B to dimuon decays in pp collisions at 7 TeV'', CERN-PH-EP-2011-120,CMS-BPH-11-002,
  arXiv:1107.5834 [hep-ex].

\bibitem{CMS_plus_LHCb}
[CMS and LHCb Collaborations],
``Search for the rare decay \Bsmumu at the LHC'',
LHCb-CONF-2011-047, CMS PAS BPH-11-019.

\bibitem{Mahmoudi:2007vz}
  F.~Mahmoudi,
  ``SuperIso: A program for calculating the isospin asymmetry of $B \to K^* \gamma$ in the MSSM'',
  Comput.\ Phys.\ Commun.\ {\bf 178} (2008) 745
  [arXiv:0710.2067].

\bibitem{Mahmoudi:2008tp}
  F.~Mahmoudi,
  ``SuperIso v2.3: A Program for calculating flavor physics observables in Supersymmetry'',
  Comput.\ Phys.\ Commun.\ {\bf 180} (2009) 1579
  [arXiv:0808.3144],\\
  \url{http://superiso.in2p3.fr} .

\bibitem{Mahmoudi:2009zz}
  F.~Mahmoudi,
  ``SuperIso v3.0, flavor physics observables calculations: Extension to
  NMSSM'',
  Comput.\ Phys.\ Commun.\  {\bf 180} (2009) 1718.

\bibitem{Grinstein:1993ys}
  B.~Grinstein,
  ``On a Precise Calculation of ($f_{B_s}/f_B$) / ($f_{D_s}/f_D$) and Its
  Implications on the Interpretation of $B -\bar{B}$ Mixing'',
  Phys.\ Rev.\ Lett.\  {\bf 71} (1993) 3067
  [hep-ph/9308226].

\bibitem{Ligeti:2003hp}
  Z.~Ligeti,
  ``$|V_{cb}|$ and $|V_{ub}|$: Theoretical developments'',
{\it In the Proceedings of Flavor Physics and CP Violation (FPCP 2003), Paris, France, 3-6 Jun 2003, pp JEU10}
  [hep-ph/0309219];
 A.~Hocker and Z.~Ligeti, ``CP violation and the CKM matrix'', Ann. Rev. Nucl. Part. Sci. {\bf 56} (2006) 501 [hep-ph/0605217].

\bibitem{Akeroyd:2010qy}
  A.~G.~Akeroyd and F.~Mahmoudi,
  ``Measuring $V_{ub}$ and probing SUSY with double ratios of purely leptonic
  decays of $B$ and $D$ mesons'',
  JHEP {\bf 1010} (2010) 038
  [arXiv:1007.2757].

\bibitem{Ellis:2005sc}
  J.~R.~Ellis, K.~A.~Olive and V.~C.~Spanos,
  ``On the interpretation of $B_s \to \mu^{+} \mu^{-}$ in the CMSSM'',
  Phys.\ Lett.\  B {\bf 624} (2005) 47
  [hep-ph/0504196].

\bibitem{Ellis:2007ss}
  J.~R.~Ellis, S.~Heinemeyer, K.~A.~Olive and G.~Weiglein,
  ``Light Heavy MSSM Higgs Bosons at Large $\tan\beta$'',
  Phys.\ Lett.\ B {\bf 653} (2007) 292
  [arXiv:0706.0977].

\bibitem{Golowich:2011cx}
  E.~Golowich, J.~Hewett, S.~Pakvasa, A.~A.~Petrov and G.~K.~Yeghiyan,
  ``Relating $B_s$ Mixing and $B_s \to \mu^+\mu^-$ with New Physics'',
  Phys.\ Rev.\  D {\bf 83} (2011) 114017
  [arXiv:1102.0009].

\bibitem{Bobeth:2001sq}
  C.~Bobeth, T.~Ewerth, F.~Kruger and J.~Urban,
  ```Analysis of neutral Higgs-boson contributions to the decays $\bar B_s \to l^+ l^-$ and $\bar B\to K l^+ l^- $'', 
  Phys.\ Rev.\  D {\bf 64} (2001) 074014
  [hep-ph/0104284].

\bibitem{Bobeth:2001jm}
  C.~Bobeth, A.~J.~Buras, F.~Kruger and J.~Urban,
  ``QCD corrections to $\bar{B} \to X_{d,s} \nu \bar{\nu}$, $\bar{B}_{d,s} \to
  \ell^{+} \ell^{-}$, $K \to \pi \nu \bar{\nu}$ and $K_{L} \to \mu^{+} \mu^{-}$
  in the MSSM'',
  Nucl.\ Phys.\  B {\bf 630} (2002) 87
  [hep-ph/0112305].

\bibitem{Buras:2002vd}
  A.~J.~Buras, P.~H.~Chankowski, J.~Rosiek and L.~Slawianowska,
  ``$\Delta M_{d,s}, B^0_{d,s} \to \mu^{+} \mu^{-}$ and $B \to X_{s} \gamma$ in
  supersymmetry at large $\tan\beta$'',
  Nucl.\ Phys.\  B {\bf 659} (2003) 3
  [hep-ph/0210145].

\bibitem{Huang:1998vb}
  C.~-S.~Huang, W.~Liao, Q.~-S.~Yan,
  ``The Promising process to distinguish supersymmetric models with large $\tan\beta$ from the standard model: $B \to X_s \mu^+ \mu^-$'',
  Phys.\ Rev.\ D {\bf 59} (1999) 011701
  [hep-ph/9803460];
  C.~-S.~Huang, Q.~-S.~Yan,
  ``$B \to X_{s} \tau^{+} \tau^{-}$ in the flipped SU(5) model'',
  Phys.\ Lett.\ B {\bf 442} (1998) 209
  [hep-ph/9803366].

\bibitem{Gamiz:2009ku}
  E.~Gamiz, C.~T.~H.~Davies, G.~P.~Lepage, J.~Shigemitsu and M.~Wingate
                  [HPQCD Collaboration],
  ``Neutral $B$ Meson Mixing in Unquenched Lattice QCD'',
  Phys.\ Rev.\  D {\bf 80} (2009) 014503
  [arXiv:0902.1815].

\bibitem{Bernard:2009wr}
  C.~Bernard {\it et al.},
  ``$B$ and $D$ Meson Decay Constants'',
  PoS {\bf LATTICE2008} (2008) 278
  [arXiv:0904.1895].

\bibitem{Laiho:2009eu}
  J.~Laiho, E.~Lunghi and R.~S.~Van de Water,
  ``Lattice QCD inputs to the CKM unitarity triangle analysis'',
  Phys.\ Rev.\  D {\bf 81} (2010) 034503
  [arXiv:0910.2928].

\bibitem{Simone:2010zz}
  J.~Simone {\it et al.}  [Fermilab Lattice and MILC Collaborations],
  ``The decay constants $f_{D_s}$, $f_{D^+}$, $f_{B_s}$ and $f_B$ from lattice QCD'',
  PoS {\bf LATTICE2010} (2010) 317.

\bibitem{:2011gx}
   {P.~Dimopoulos \it et al.}  [ETM Collaboration],
  ``Lattice QCD determination of $m_b$, $f_B$ and $f_{B_s}$ with twisted mass Wilson
  fermions'',
  arXiv:1107.1441 [hep-lat].

\bibitem{Lattice2011}
Lattice 2011, Lake Tahoe, California, USA, 11-16 July 2011.

\bibitem{Shigemitsu:2011sp}
  J.~Shigemitsu, H.~Na, C.~Davies, R.~Horgan, C.~Monahan and P.~Lepage,
  arXiv:1110.5783 [hep-lat].


\bibitem{McNeile:2011ng}
  C.~McNeile, C.~T.~H.~Davies, E.~Follana, K.~Hornbostel and G.~P.~Lepage,
  arXiv:1110.4510 [hep-lat].



\bibitem{Oakes:1994tj}
  R.~J.~Oakes,
  ``Ratios of charmed and beauty meson decay constants'',
  Phys.\ Rev.\ Lett.\  {\bf 73} (1994) 381.

\bibitem{Bona:2009cj}
  M.~Bona {\it et al.}  [UTfit Collaboration],
  ``An Improved Standard Model Prediction Of BR($B \to \tau \nu$) And Its
  Implications For New Physics'',
  Phys.\ Lett.\  B {\bf 687} (2010) 61
  [arXiv:0908.3470].

\bibitem{Asner:2010qj}
  D.~Asner {\it et al.}  [Heavy Flavor Averaging Group],
  ``Averages of $b$-hadron, $c$-hadron, and $\tau$-lepton Properties'',
  arXiv:1010.1589 [hep-ex].

\bibitem{Ha:2010rf}
  H.~Ha {\it et al.}  [BELLE Collaboration],
  ``Measurement of the decay $B^0\to\pi^-\ell^+\nu$ and determination of
  $|V_{ub}|$'',
  Phys.\ Rev.\  D {\bf 83} (2011) 071101
  [arXiv:1012.0090].

\bibitem{:2010uj}
  P.~del Amo Sanchez {\it et al.}  [BABAR Collaboration],
  ``Study of $B \to \pi \ell \nu$ and $B \to \rho \ell \nu$ Decays and
  Determination of $|V_{ub}|$'',
  Phys.\ Rev.\  D {\bf 83} (2011) 032007
  [arXiv:1005.3288].

\bibitem{Sigamani:2011ne}
  M.~Sigamani  [On behalf of the BABAR collaboration],
  ``Measurements of the Partial Branching Fraction for $B\to X_u \ell \nu$ and the
  Determination of $V_{ub}$'',
  PoS {\bf ICHEP2010} (2010) 265
  [arXiv:1103.0560].

\bibitem{:2009tp}
  P.~Urquijo {\it et al.}  [Belle Collaboration],
  ``Measurement Of $|V_{ub}|$ From Inclusive Charmless Semileptonic $B$ Decays'',
  Phys.\ Rev.\ Lett.\  {\bf 104} (2010) 021801
  [arXiv:0907.0379].

\bibitem{Akeroyd:2009tn}
  A.~G.~Akeroyd and F.~Mahmoudi,
  ``Constraints on charged Higgs bosons from $D_s^{\pm} \to \mu^{\pm} \nu$ and $D_s^{\pm} \to \tau^{\pm} \nu$'',
  JHEP {\bf 0904} (2009) 121
  [arXiv:0902.2393].

 \bibitem{Hou:1992sy}
   W.~S.~Hou,
   ``Enhanced charged Higgs boson effects in $B^- \to \tau \bar{\nu}, \mu \bar{\nu}$ and $b \to \tau \bar{\nu} + X$'',
   Phys.\ Rev.\ D {\bf 48} (1993) 2342.
 
\bibitem{Akeroyd:2003zr}
   A.~G.~Akeroyd and S.~Recksiegel,
   ``The effect of $H^\pm$ on $B^\pm \to \tau^\pm \nu_\tau$ and  $B^\pm \to \mu^\pm \nu_\mu$'',
   J.\ Phys.\ G {\bf 29} (2003) 2311
   [hep-ph/0306037].

\bibitem{Itoh:2004ye}
  H.~Itoh, S.~Komine and Y.~Okada,
  ``Tauonic $B$ decays in the minimal supersymmetric standard model'',
  Prog.\ Theor.\ Phys.\  {\bf 114} (2005) 179
  [hep-ph/0409228].

\bibitem{Isidori:2006pk}
  G.~Isidori and P.~Paradisi,
  ``Hints of large $\tan\beta$ in flavour physics'',
  Phys.\ Lett.\  B {\bf 639} (2006) 499
  [hep-ph/0605012].

\bibitem{Asner:2008nq}
  D.~M.~Asner {\it et al.},
   ``Physics at BES-III'',
  Int. J. Mod. Phys. A {\bf 24}, suppl. 1 (2009)
  [arXiv:0809.1869].

\bibitem{Farina:2011bh}
  M.~Farina, M.~Kadastik, D.~Pappadopulo, J.~Pata, M.~Raidal and A.~Strumia,
  ``Implications of XENON100 and LHC results for Dark Matter models'', 
  Nucl.\ Phys.\ B {\bf 853} (2011) 607 
  [arXiv:1104.3572].

\bibitem{Dutta:2011bk}
  B.~Dutta, Y.~Mimura and Y.~Santoso,
  ``$B_s \to \mu^+\mu^-$ in Supersymmetric Grand Unified Theories'',
  arXiv:1107.3020 [hep-ph];
  S.~Akula, D.~Feldman, P.~Nath and G.~Peim,
 ``Excess Observed in CDF $B^0_s \to \mu^{+} \mu^{-}$ and SUSY at the LHC'',
  Phys.\ Rev.\ D {\bf 84} (2011) 115011 
  [arXiv:1107.3535];
  D.~Hooper and C.~Kelso,
  ``Implications of a Large $B_s \rightarrow \mu^+ \mu^-$ Branching Fraction
  for the Minimal Supersymmetric Standard Model'',
  arXiv:1107.3858 [hep-ph];
  W.~Altmannshofer, M.~Carena, S.~Gori and A.~de la Puente,
  ``Signals of CP Violation Beyond the MSSM in Higgs and Flavor Physics'',
  Phys.\ Rev.\ D {\bf 84} (2011) 095027
  [arXiv:1107.3814].


\bibitem{Allanach:2001kg}
  B.~C.~Allanach,
  ``SOFTSUSY: a program for calculating supersymmetric spectra'',
  Comput.\ Phys.\ Commun.\  {\bf 143} (2002) 305
  [hep-ph/0104145].

\bibitem{Ellwanger:2006rn}
  U.~Ellwanger and C.~Hugonie,
  ``NMSPEC: A Fortran code for the sparticle and Higgs masses in the NMSSM with
  GUT scale boundary conditions'',
  Comput.\ Phys.\ Commun.\  {\bf 177} (2007) 399
  [hep-ph/0612134].

\bibitem{Bechtle:2008jh}
  P.~Bechtle, O.~Brein, S.~Heinemeyer, G.~Weiglein and K.~E.~Williams,
  ``HiggsBounds: Confronting Arbitrary Higgs Sectors with Exclusion Bounds from
  LEP and the Tevatron'',
  Comput.\ Phys.\ Commun.\  {\bf 181} (2010) 138
  [arXiv:0811.4169].

\bibitem{Bechtle:2011sb}
  P.~Bechtle, O.~Brein, S.~Heinemeyer, G.~Weiglein and K.~E.~Williams,
  ``HiggsBounds 2.0.0: Confronting Neutral and Charged Higgs Sector Predictions
  with Exclusion Bounds from LEP and the Tevatron'',
  Comput.\ Phys.\ Commun.\ {\bf 182} (2011) 2605
  [arXiv:1102.1898].

\bibitem{:1900yx}
   [Tevatron Electroweak Working Group for CDF and D0 Collaboration],
  ``Combination of CDF and D0 Results on the Mass of the Top Quark'',
  arXiv:1007.3178 [hep-ex].

\bibitem{mSUGRA}
  A.~H.~Hamseddine, R.~Arnowitt and P.~Nath,
  ``Localy Supersymmetric Grand Unification'',
  Phys.\ Rev.\ Lett.\ {\bf 49} (1982) 970;
%
  R.~Baieri, S.~Ferrara and C.~A.~Savoy,
  ``Gauge Models With Spontaneously Broken Local Supersymmetry'',
  Phys.\ Lett.\ B {\bf 119} (1982) 343;
%
  L.~J.~Hall, J.~D.~Lykken and S.~Weinberg,
  ``Supergravity As The Messenger Of Supersymmetry Breaking'',
  Phys.\ Rev.\ D {\bf 27} (1983) 2359;
%
  N.~Ohta,
  ``Grand Unified Theories Based On Local Supersymmetry'',
  Prog.\ Theor.\ Phys.\ {\bf 70} (1983) 542.

\bibitem{atlas-cms-EPS}
As presented at the EPS 2011 conference.

\bibitem{Ellis:2002wv}
  J.~R.~Ellis, K.~A.~Olive and Y.~Santoso,
  ``The MSSM Parameter Space with Non-Universal Higgs Masses'',
  Phys.\ Lett.\ B {\bf 539}, 107 (2002)
  [hep-ph/0204192].

\bibitem{AMSB}
  L.~Randall and R.~Sundrum,
  ``Out of this world supersymmetry breaking'',
  Nucl.\ Phys.\  B {\bf 557} (1999) 79
  [hep-th/9810155];
%
  G.~F.~Giudice, M.~A.~Luty, H.~Murayama and R.~Rattazzi,
  ``Gaugino mass without singlets'',
  JHEP {\bf 9812} (1998) 027
  [hep-ph/9810442];
%
  T.~Gherghetta, G.~F.~Giudice and J.~D.~Wells,
  ``Phenomenological consequences of supersymmetry with anomaly induced
  masses'',
  Nucl.\ Phys.\  B {\bf 559} (1999) 27
  [hep-ph/9904378].

\bibitem{Arbey:2011gu}
  A.~Arbey, A.~Deandrea and A.~Tarhini,
  ``Anomaly mediated SUSY breaking scenarios in the light of cosmology and in
  the dark (matter)'',
  JHEP {\bf 1105} (2011) 078
  [arXiv:1103.3244].

\bibitem{GMSB}
  M.~Dine, W.~Fischler and M.~Srednicki,
  ``Supersymmetric Technicolor'',
  Nucl.\ Phys.\  B {\bf 189} (1981) 575;
%
  S.~Dimopoulos and S.~Raby,
  ``Supercolor'',
  Nucl.\ Phys.\  B {\bf 192} (1981) 353;
%
  L.~Alvarez-Gaume, M.~Claudson and M.~B.~Wise,
  ``Low-Energy Supersymmetry'',
  Nucl.\ Phys.\  B {\bf 207} (1982) 96;
%
  C.~R.~Nappi and B.~A.~Ovrut,
  ``Supersymmetric Extension of the SU(3) x SU(2) x U(1) Model'',
  Phys.\ Lett.\  B {\bf 113} (1982) 175;
%
  M.~Dine, A.~E.~Nelson, Y.~Nir and Y.~Shirman,
  ``New tools for low-energy dynamical supersymmetry breaking'',
  Phys.\ Rev.\  D {\bf 53} (1996) 2658
  [hep-ph/9507378].

\bibitem{NMSSM}
  T.~Elliott, S.~F.~King and P.~L.~White,
  ``Unification constraints in the next-to-minimal supersymmetric standard
  model'',
  Phys.\ Lett.\  B {\bf 351} (1995) 213
  [hep-ph/9406303];
%
  S.~F.~King and P.~L.~White,
  ``Resolving the constrained minimal and next-to-minimal supersymmetric
  standard models'',
  Phys.\ Rev.\  D {\bf 52} (1995) 4183
  [hep-ph/9505326];
%
  U.~Ellwanger, M.~Rausch de Traubenberg and C.~A.~Savoy,
  ``Phenomenology of supersymmetric models with a singlet'',
  Nucl.\ Phys.\  B {\bf 492} (1997) 21
  [hep-ph/9611251].


\bibitem{Belanger:2008nt}
  G.~B\'elanger, C.~Hugonie and A.~Pukhov,
  ``Precision measurements, dark matter direct detection and LHC Higgs searches
  in a constrained NMSSM'',
  JCAP {\bf 0901} (2009) 023
  [arXiv:0811.3224].

\bibitem{Domingo:2007dx}
 F.~Domingo and U.~Ellwanger,
 ``Updated Constraints from $B$ Physics on the MSSM and the NMSSM'',
 JHEP {\bf 0712} (2007) 090
 [arXiv:0710.3714].

\bibitem{Mahmoudi:2010xp}
  F.~Mahmoudi, J.~Rathsman, O.~St{\aa}l and L.~Zeune,
  ``Light Higgs bosons in phenomenological NMSSM,''
  Eur.\ Phys.\ J.\  C {\bf 71} (2011) 1608
  [arXiv:1012.4490].

\bibitem{mc_limit}
   T. Junk,
  ``Sensitivity, Exclusion and Discovery with Small Signals, Large Backgrounds, and Large Systematic Uncertainties'',
  CDF/DOC/STATISTICS/PUBLIC/8128.
 
\bibitem{fdfs}
  R.~Aaij {\it et al.} [LHCb Collaboration], `` Average $f_s/f_d$ $b$-hadron production fraction for 7 TeV $pp$ collision'', LHCb-CONF-2011-034, 26 July 2011.

\bibitem{thesis_2010_068} D. Mart{\'{\i}}nez Santos, ``Study of the very rare decay \Bsmumu in LHCb'', CERN-THESIS-2010-068.

\bibitem{Dedes:2008iw}
  A.~Dedes, J.~Rosiek and P.~Tanedo,
  ``Complete One-Loop MSSM Predictions for $B^0 \to \ell^+ \ell^{\prime-}$ at the
  Tevatron and LHC'',
  Phys.\ Rev.\  D {\bf 79} (2009) 055006
  [arXiv:0812.4320].


\end{thebibliography}
\end{document}